\documentclass[twocolumn,trackchanges,times]{aastex631}

\usepackage{CJK}
\usepackage{xspace} 
\usepackage{amsmath}
\usepackage{url}
\usepackage{natbib}
\usepackage{tabularx}
\usepackage{color}
\usepackage{graphicx}
\usepackage{epstopdf}
\usepackage{gensymb}
\usepackage{hyperref}
\usepackage{longtable}
\usepackage{subfigure}
\usepackage[T1]{fontenc}
\DeclareUnicodeCharacter{2212}{-}

\newcommand*{\ac}{$A$(C)\xspace}

\newcommand*{\teff}{$T_{\rm eff}$\xspace}
\newcommand*{\logg}{$\log~g$\xspace}
\newcommand*{\feh}{[Fe/H]\xspace}

\newcommand*{\cfe}{[C/Fe]\xspace}

\newcommand*{\kms}{km s$^{-1}$\xspace}

\newcommand*{\msun}{$M_\odot$\xspace}

\newcommand*{\cemps}{CEMP-$s$\xspace}

\newcommand*{\cempno}{CEMP-$\rm no$}\xspace

\newcommand*{\gaia}{$\it Gaia$\xspace}

\begin{document}

\shorttitle{A New Subclass of CEMP Stars}
\shortauthors{Lee et al.}
\begin{CJK}{UTF8}{mj}
\title{A New Subclass of Carbon-Enhanced Metal-Poor Stars at Extremely Low Metallicity}

\author[0000-0001-5297-4518]{Young Sun Lee (이영선)}
\affiliation{Department of Astronomy and Space Science, Chungnam National University, Daejeon 34134, Republic of Korea; youngsun@cnu.ac.kr}

\author[0000-0003-4573-6233]{Timothy C. Beers}
\affiliation{Department of Physics and Astronomy, University of Notre Dame, Notre Dame, IN 46556, USA}
\affiliation{Joint Institute for Nuclear Astrophysics -- Center for the Evolution of the Elements (JINA-CEE), USA}

\author[0000-0002-5661-033X]{Yutaka Hirai}
\affiliation{Department of Community Service and Science, Tohoku University of Community Service and Science, 3-5-1 Iimoriyama, Sakata, Yamagata 998-8580, Japan}

\author[0000-0002-2453-0853]{Jihye Hong}
\affiliation{Department of Physics and Astronomy, University of Notre Dame, Notre Dame, IN 46556, USA}
\affiliation{Joint Institute for Nuclear Astrophysics -- Center for the Evolution of the Elements (JINA-CEE), USA}

\author[0009-0009-7838-7771]{Miji Jeong}
\affiliation{Department of Astronomy, Space Science, and Geology, Chungnam National University, Daejeon 34134, Republic of Korea}
\affiliation{Korea Astronomy and Space Science Institute, Daejeon 34055, Republic of Korea}

\author[0009-0000-8578-1317]{Changmin Kim}
\affiliation{Department of Earth, Environmental \& Space Sciences, Chungnam National University, Daejeon 34134, Republic of Korea}

\author[0000-0002-6411-5857]{Young Kwang Kim}
\affiliation{Department of Astronomy and Space Science, Chungnam National University, Daejeon 34134, Republic of Korea}

\begin{abstract}

We report the discovery of a new subclass of carbon-enhanced metal-poor (CEMP) stars, characterized by high absolute carbon abundances (\ac $>$ 7.39) and extremely low metallicity ([Fe/H] $\leq$ --3.1) but notably lacking enhancements in neutron-capture elements, thus falling under the CEMP-no category. This population emerged from a detailed analysis of low-resolution spectroscopic data obtained from the Sloan Digital Sky Survey (SDSS) and the Large Sky Area Multi-Object Fiber Spectroscopic Telescope (LAMOST), where the observed frequency trends with the decreasing metallicity of \cemps ($s$-process-enhanced) and CEMP-no (no neutron-capture enhanced) stars deviated from established expectations. In contrast to earlier findings, we observe a rise in high-\ac stars below [Fe/H] = --3.1, which we interpret as a distinct group not accounted for in traditional CEMP classifications. Following the Yoon-Beers group classification, we define these stars as Group IV. Statistical modeling confirms their presence as a separate peak in the \ac distribution, and available radial velocity data suggest that about 30\% of Group IV stars may be binaries, indicating possible binary-related formation mechanisms. This discovery challenges the current CEMP-no star formation pathways and implies the existence of alternative or hybrid enrichment scenarios in the early Universe. High-resolution spectroscopic follow-up of Group IV candidates will be crucial for identifying their progenitors and understanding their evolutionary implications.

\end{abstract}


\section {Introduction} \label{sec:intro}

Detailed chemical-abundance studies of very metal-poor (VMP; [Fe/H]\footnote[9]{[Fe/H] = log$_{10}\,$($N$(Fe)/$N$(H))$_\star$ -- log$_{10}\,$($N$(Fe)/$N$(H))$_\odot$, where $N$(Fe) and $N$(H) represent the number densities of iron and hydrogen, respectively.} $\leq$ --2.0) stars in the halo system of the Milky Way (MW) have shown that, while most exhibit similar abundance patterns, there are numerous cases of stars with peculiar chemical compositions, including significantly high or low abundances in light elements such as C, N, O, Na, Mg, Si, and others \citep[e.g.,][]{norris2013,li2022,jeong2023}. The most common class among such stars is that of objects with elevated carbon levels, referred to as carbon-enhanced metal-poor (CEMP) stars, originally defined as having a metallicity of [Fe/H] $\leq$ --1.0 and [C/Fe] $> +1.0$ \citep{beers2005}. In recent years, most studies have adopted a definition of CEMP stars with [C/Fe] $> +0.7$ \citep[e.g.,][]{aoki2007,carollo2012,lee2013,lee2017,lee2019,arentsen2023,lucey2023,arentsen2025}.

One intriguing aspect of CEMP stars is the notable rise in their proportions, compared to ``normal'' iron-poor stars, as metallicity decreases. The cumulative fraction of CEMP stars increases dramatically, beginning at roughly 20\% for [Fe/H] $<$ --2.0, rising to $\sim 30$\% for [Fe/H] $<$ --3.0, reaching $\sim 40$\% for [Fe/H] $<$ --3.5, and culminating at over $\sim 80$\% for [Fe/H] $<$ --4.0 \citep[e.g.,][]{lucatello2006,lee2013,yong2013,placco2014,yoon2018,arentsen2022}.

Subsequent examination of the chemical makeup of CEMP stars has resulted in their division into various groups based on enrichment in neutron-capture ($n$-capture) elements: CEMP-no, CEMP-$s$, CEMP-$r$, and CEMP-$r/s$ \citep{beers2005}. CEMP-no stars show no excess of heavy $n$-capture elements, while CEMP-$s$ stars exhibit notable overabundances of $s$-process elements such as Ba. CEMP-$r$ stars are distinguished by high abundances of $r$-process elements like Eu. Stars exhibiting possible contributions from both the $r$-process and $s$-process are classified as CEMP-$r/s$, which is likely linked to an intermediate $n$-capture process ($i$-process; see \citealt{cowan1977,hampel2016,frebel2018,choplin2024} and references therein).

CEMP-no and \cemps\ stars represent the two largest subclasses within the CEMP classification, accounting for more than 95\% of all CEMP stars. CEMP-no stars appear to be the majority population of CEMP stars with [Fe/H] $<$ --3.0, whereas \cemps\ stars are predominantly found in [Fe/H] $>$ --3.0 (e.g., \citealt{aoki2007,yong2013,yoon2016}). Comprehensive radial-velocity (RV) analyses of CEMP stars have yielded strong evidence for differing binary frequencies between \cemps\ and CEMP-no stars (e.g., \citealt{starkenburg2014,hansen2016a,hansen2016b,jorissen2016}). Roughly 82\% of \cemps\ stars (and CEMP-$r/s$ stars) are found in binary systems. In contrast, only about 17\% of CEMP-no stars appear to be members of binary systems, similar to the observed binary rate among other metal-poor halo stars \citep{carney2003,starkenburg2014,hansen2016a}. This fraction was higher (32\%) in the work of \cite{arentsen2019}, who confirmed four additional CEMP-no binaries based on RV monitoring.

The diverse categorizations and varying binary status of CEMP stars require multiple astrophysical sources for carbon creation. Proposed mechanisms include: (1) for \cemps\ (and CEMP-$r/s$), the transfer of carbon-rich material from a companion asymptotic giant branch (AGB) star to the presently observed star \citep[e.g.,][]{suda2004,herwig2005,sneden2008,masseron2010,bisterzo2011,bisterzo2012}; and (2) for CEMP-no, enrichment by massive, rapidly rotating stars with zero (or extremely low) metallicity, which produce considerable amounts of C, N, and O through distinctive internal burning and mixing processes \citep{meynet2006,meynet2010,choplin2016,choplin2017,liu2021}, or by faint supernovae (SNe) associated with the earliest generations of stars, which undergo extensive mixing and fallback during their explosions, releasing large quantities of C and O \citep{umeda2003,umeda2005,tominaga2007,tominaga2014,ito2009,ito2013}.

Given the differences in the observed properties and proposed progenitors of \cemps\ and CEMP-no stars, determining their comparative frequencies across the stellar populations of the MW is essential to constrain their genesis and associated chemical-enrichment processes. A key challenge in this pursuit is assembling a sufficiently large and homogeneous sample of CEMP stars. Low-resolution spectroscopic data have proven valuable in this regard, enabling the determination of CEMP star frequency over a wide range of metallicities \citep[e.g.,][]{lee2013,placco2014,yoon2018,arentsen2021,lucey2023,arentsen2025} and exploration of their frequency across MW components such as the bulge \citep{howes2016,arentsen2021}, the disk system \citep{dietz2021}, the halo system \citep{frebel2006,carollo2012,carollo2014,lee2017,yoon2018,lee2019}, and satellite dwarf galaxies \citep[e.g.,][]{starkenburg2013,salvadori2015,chiti2018,yoon2019,ji2020}.

Another hurdle is the need to confidently differentiate \cemps\ stars from CEMP-no stars. \cite{yoon2016} demonstrated that the previously recognized classes of CEMP-no stars (their Groups~II and III) can be usefully separated from CEMP-$s$ and CEMP-$r/s$ stars (their Group~I) using the absolute carbon abundance, \ac\footnote[10]{The conventional notation is used, \ac\ $=$ log\,$\epsilon$(C) $=$ log$_{10}$\,($N_{\rm C}$/$N_{\rm H}$) + 12, where $N_{\rm C}$ and $N_{\rm H}$ represent the number densities of carbon and hydrogen, respectively.}, which can be obtained from low-resolution spectra.

Here, we employ the extensive set of low-resolution stellar spectra from the Sloan Digital Sky Survey (SDSS; \citealt{york2000}) and the Large Sky Area Multi-Object Fiber Spectroscopic Telescope (LAMOST; \citealt{cui2012,luo2015}). Using the CEMP-no and \cemps\ separation method proposed by \cite{yoon2016}, we investigate the relative proportions of CEMP-no and \cemps\ stars as a function of [Fe/H]. While constructing the CEMP frequencies versus [Fe/H] for the two subclasses, we notice an unexpected behavior: the cumulative frequency of CEMP-no stars (classified as such based on their low \ac) \emph{does not} increase substantially for [Fe/H] $<$ --2.5, but instead behaves similarly to that of the \cemps\ stars (classified as such based on their high \ac), which rises to 30\% for [Fe/H] $<$ --4.0. This behavior is opposite to the findings of \cite{yoon2018} and to trends based on high-resolution spectroscopy. Using low-resolution spectra from the AAOmega Evolution of Galactic Structure (AEGIS) survey (PI: Keller), \cite{yoon2018} reported a flat cumulative fraction of \cemps\ stars for [Fe/H] $<$ --2.5, while the CEMP-no fraction continues to increase with declining [Fe/H]. The high-resolution spectroscopic studies also exhibit a continuous rise of the CEMP-no star fraction with decreasing metallicity. To address this apparent contradiction, we propose the existence of a new subclass of high-\ac, CEMP-no stars with [Fe/H] $\leq$ --3.1 and \ac\ $>$ 7.39, which we designate, following the scheme of \cite{yoon2016}, as Group~IV.

The structure of this paper is as follows. Section~\ref{sec2} outlines our sample selection from SDSS and LAMOST and the method for merging these datasets. Section~\ref{sec3} describes our adopted technique for distinguishing CEMP-no from \cemps\ stars. In Section~\ref{sec4}, we derive and analyze the cumulative frequencies of CEMP, \cemps, and CEMP-no stars as a function of [Fe/H], and propose the new subclass of CEMP-no stars. And then we explore the properties of the proposed new subclass of CEMP-no stars. Section~\ref{sec5} discusses potential progenitors of CEMP-no stars and presents our conclusions.


\begin{figure} [t]
\begin{center}
\includegraphics[width=\columnwidth]{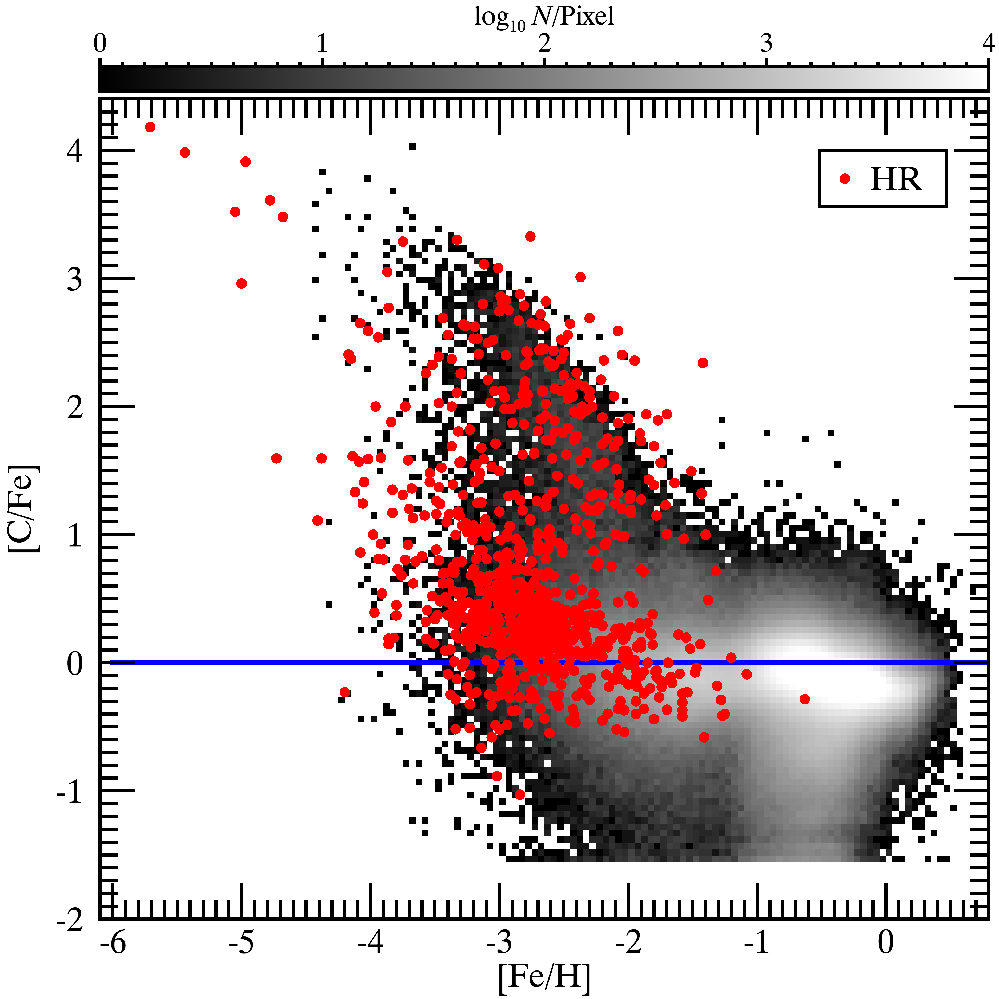}
\caption{Logarithmic number-density plot of our final SDSS/LAMOST sample in the [C/Fe]--[Fe/H] plane, after smoothing with a Gaussian kernel, coded as shown in the bar at the top. The blue-solid line is at [C/Fe] = 0. The red dots are taken from the high-resolution spectroscopic determinations in various literature (see text). Each bin has a size of 0.05$\times$0.05 dex in [Fe/H] and [C/Fe], respectively.}
\label{fig1}
\end{center}
\end{figure}

\section{The SDSS and LAMOST Samples} \label{sec2}

This study employs data from two major spectroscopic surveys. The first is the original Sloan Digital Sky Survey (SDSS) and its subsequent iterations, including the Sloan Extension for Galactic Understanding and Exploration (SEGUE; \citealt{yanny2009,rockosi2022}), the Baryon Oscillation Spectroscopic Survey (BOSS; \citealt{dawson2013}), and the extended BOSS (eBOSS; \citealt{blanton2017}). Low-resolution ($R\sim1800$) stellar spectra from these surveys were processed using an updated version of the SEGUE Stellar Parameter Pipeline (SSPP; \citealt{allende2008,lee2008a,lee2008b,lee2011,smolinski2011,lee2013}), which provides precise estimates of stellar atmospheric parameters -- effective temperature (\teff), surface gravity (\logg), metallicity (\feh) -- along with [C/Fe]. Typical $1\sigma$ uncertainties are 180 K for \teff, 0.24 dex for \logg, 0.23 dex for \feh, and less than 0.3 dex for [C/Fe].

The second source is the LAMOST Data Release 6 (http://www.lamost.org/lmusers/), which contributes over six million stellar spectra. Given its similar wavelength coverage (3800--9000 \AA) and resolution ($R\sim1800$), the SSPP can be applied directly to LAMOST spectra to obtain the same set of parameters, including [C/Fe] ratios (see \citealt{lee2015} for details).

When combining stellar samples from different surveys, it is critical to place their parameter and abundance scales onto a common system \citep[e.g.,][]{arentsen2022}. Following \citet{lee2023}, we quantified systematic differences using $\sim$ 44,000 stars in common between SDSS and LAMOST. Average offsets are minimal -- 5 K in \teff, 0.04 dex in \logg, 0.10 dex in [Fe/H], and less than 0.02 dex in [C/Fe] -- well below the individual measurement uncertainties. Accordingly, no corrections for systematic offsets were applied.

To assemble a high-quality dataset for studying CEMP stars, we excluded stars in the direction of known open and globular clusters, retained only the highest signal-to-noise ratio (S/N) spectrum for multiply observed stars, and imposed conservative cuts: S/N $\geq$ 15 \AA$^{-1}$, 4000 $\leq$ \teff $\leq$ 6700, and --4.5 $\leq$ \feh $\leq$ +0.5. We then visually inspected spectra of stars with [Fe/H] $\leq -2.0$ to remove cool white dwarfs and objects with Ca {\sc ii} emission cores that can bias \feh\ estimates. Additionally, for stars with [Fe/H] $>$ --2.0 but SSPP-derived [C/Fe] $>$ +0.7, we inspected the spectra to discard unreliable measurements. After these steps, our final SDSS and LAMOST sample (hereafter, LR sample) contains $\sim$3.15 million stars -- the largest spectroscopic dataset to date with uniformly measured [C/Fe].

Because our primary interest lies in the low-metallicity regime ([Fe/H] $<$ --2.0), we evaluated the reliability of SSPP-derived [C/Fe] in this domain. We compiled a high-resolution comparison sample (hereafter, HR sample) from \citet{yoon2016}, \citet{arentsen2019}, and the Stellar Abundances for Galactic Archaeology (SAGA) database \citep{suda2008}, then cross-matched it with the LR sample. This yielded 92 stars in common, for which we compared SSPP-derived [Fe/H] and [C/Fe] to HR values. We found small zero-point offsets of 0.13 dex ($\sigma$ = 0.29 dex) for [Fe/H] and 0.13 dex ($\sigma$ = 0.30 dex) for [C/Fe], with no significant trends. Given these modest differences, we applied the offsets to the LR sample to place both datasets on a common abundance scale. Duplicate entries within the HR sample were removed, and for overlapping stars between the HR and LR samples, we adopted the HR measurements in the subsequent analyses.

Figure~\ref{fig1} shows the logarithmic number-density distribution of our full sample in the [C/Fe]--[Fe/H] plane, smoothed with a Gaussian kernel (pixel size: 0.05 dex $\times$ 0.05 dex). As metallicity decreases, the [C/Fe] distribution broadens for stars with \feh\ $<$ --1.5 and [C/Fe] $>$ 0.0, indicating an increasing fraction of carbon-rich stars among the metal-poor population. A striking rise in stars with [C/Fe] $>$ +2.0 occurs below [Fe/H] = --2.0, consistent with previous findings \citep[e.g.,][]{lee2013,placco2014,yoon2019,arentsen2025}. At a given metallicity below this threshold, the [C/Fe] distribution exhibits an extended high-abundance tail. The red dots in the figure indicate measurements from the HR sample.

\begin{figure} [!t]
\begin{center}
\includegraphics[width=\columnwidth]{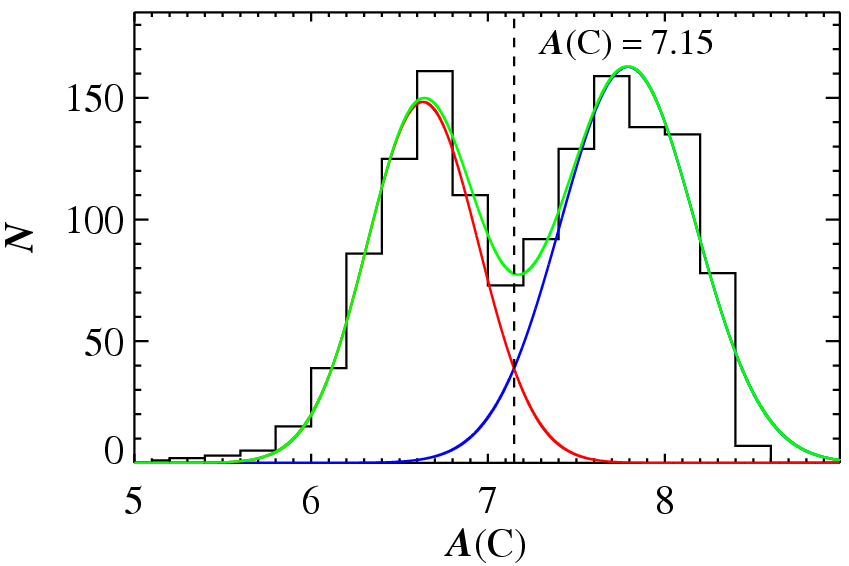}
\caption{Distribution of \ac\ for the high-quality spectra (S/N $\geq$ 30) of stars with [Fe/H] $<$ --2.5. The red and blue curves are fits to a Gaussian mixture model for the low and high-\ac stars, respectively. The green curve is the sum of the two Gaussian curves. The vertical dashed line at \ac = 7.15 separates \cempno\ stars from \cemps\ stars. }
\label{fig2}
\end{center}
\end{figure}

\section{Separation of CEMP-no stars From CEMP-$s$ Stars} \label{sec3}

We aim to derive the fractions of CEMP-no and \cemps\ stars, and to examine whether an additional subclass of CEMP-no stars might exist, based on the derived fractions. To do so, it is necessary to establish clear criteria for distinguishing between CEMP-no and \cemps\ stars. Traditionally, this classification relies on [Ba/Fe] abundance ratios obtained from high-resolution spectroscopy. However, for our low-resolution sample, this approach is not readily applicable due to the intrinsically weak absorption features of $n$-capture elements.

Fortunately, the differences between the CEMP subclasses are not limited to their $n$-capture abundances. They also exhibit distinct patterns in the \ac--[Fe/H] plane, often showing a bimodal distribution \citep[e.g.,][]{rossi2005,spite2013,bonifacio2015,yoon2016,lee2017,lee2019}. \citet{yoon2016} demonstrated that these subclasses can be reliably separated using \ac, which can be measured from low-resolution spectra. We therefore adopt this \ac-based method to subclassify our program stars.

Figure \ref{fig2} illustrates this approach. The histogram clearly exhibits a bimodal distribution in \ac. To distinguish the low-\ac\ stars (classified as CEMP-no) from the high-\ac\ stars (classified as \cemps), we fit two Gaussian curves to the \ac\ distribution. The intersection point of the two Gaussians occurs at \ac\ = 7.15, indicated by the black dashed line in the figure. We classify stars with \ac\ $\le$ 7.15 as CEMP-no and those with \ac\ $>$ 7.15 as \cemps. Note that Figure \ref{fig2} includes only stars with [Fe/H] $<$ --2.5 and [C/Fe] $>$ +0.7, and with spectra having S/N $\geq$ 30, in order to establish a clear and robust boundary between the subclasses.

The \ac-based classification is not entirely free from the possibility of misidentification. \citet{yoon2016} estimated that the misclassification rate is below 10\% when compared with the [Ba/Fe]-based classification, which is sufficiently low for our purposes. In addition, some overlap can occur between CEMP-no and CEMP-$r$, as well as between CEMP-$s$ and CEMP-$r/s$ (CEMP-$i$) stars. However, since CEMP-$r$ and CEMP-$i$ stars represent only a small fraction of the overall CEMP population, their potential contamination is expected to be negligible.

\begin{figure}[t]
\centering
\includegraphics[width=\columnwidth]{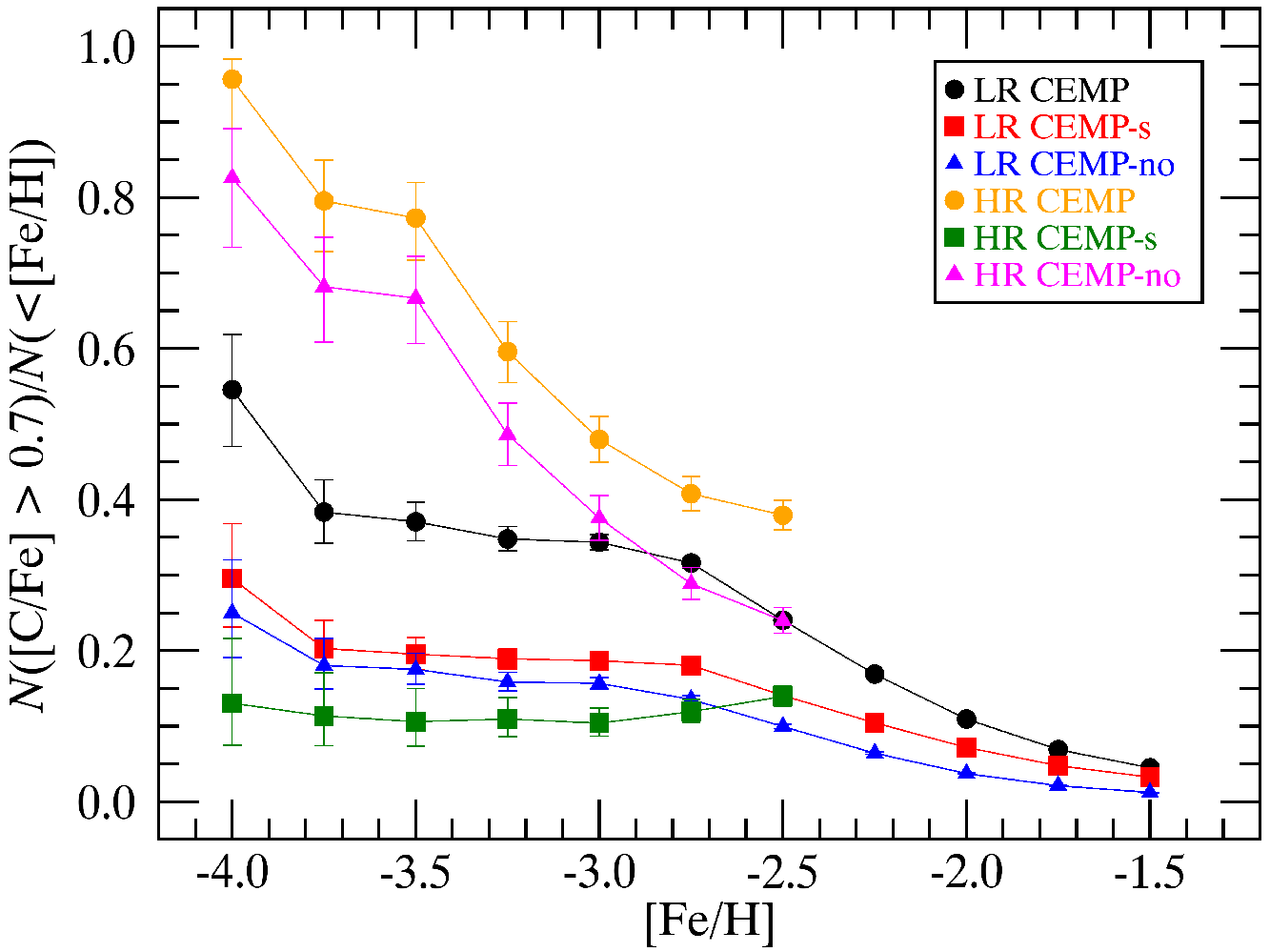}
\caption{Cumulative frequencies, as a function of [Fe/H], for our full sample of CEMP (black), \cemps\ (red), and \cempno\ (blue) stars, along with their associated error bars. The orange, green, and magenta symbols represent the frequency of each category from the HR sample. The HR sample includes stars with [Fe/H] $<$ --2.5. The error bars are the 1$\sigma$ binomial Wilson score confidence intervals (see text).}
\label{fig3}
\end{figure}

\section{Results}\label{sec4}
\subsection{Frequencies of CEMP Stars} \label{sec41}

We now derive the frequencies of CEMP stars from our large LR sample. Owing to the low resolution of the SDSS/LAMOST spectra, our \cfe\ measurements are less precise than those from high-resolution analyses; however, the much larger sample size relative to previous work enables us to identify meaningful trends in CEMP frequency. For clearly detected C-enhanced stars, we count as C-rich stars only those with \cfe\ $> +0.7$ and a correlation coefficient (CC) $>$ 0.5. The CC is computed by comparing each observed spectrum with a synthetic spectrum (using SSPP parameters and [C/Fe]) over the CH $G$-band (4290--4318\,\AA). The cumulative frequency of C-enhanced stars is then obtained by dividing the number of C-rich stars by all stars below a given metallicity, following Equation (3) of \citet{lee2013}. Because C-rich stars with CC $<$ 0.5 are excluded -- either due to poor spectra or poor matches to the synthetic spectra -- our CEMP fractions should be regarded as lower limits.

Figure~\ref{fig3} shows the derived cumulative frequencies of CEMP (black), \cemps\ (red), and \cempno\ (blue) stars, with associated error bars, as a function of [Fe/H]. Because the CEMP fraction is very low at [Fe/H] $>$ --1.5, we confine the metallicity range to [Fe/H] $<$ --1.5. Error bars are 68\% Wilson score confidence intervals for binomial proportions \citep{wilson1927}, which provide more realistic uncertainties for smaller samples (particularly at [Fe/H] $<$ --3.0). As described above, we separate CEMP-no and \cemps\ stars at \ac\ = 7.15 (CEMP-no: \ac\ $\le$ 7.15; \cemps: \ac\ $>$ 7.15). For comparison, we also include the HR-sample frequencies -- orange (CEMP), green (\cemps), and magenta (CEMP-no) -- considering only [Fe/H] $<$ --2.5, since the HR sample at higher metallicity is biased toward CEMP stars.

\begin{figure} [!t]
\centering
\epsscale{1.11}
\includegraphics[width=\columnwidth]{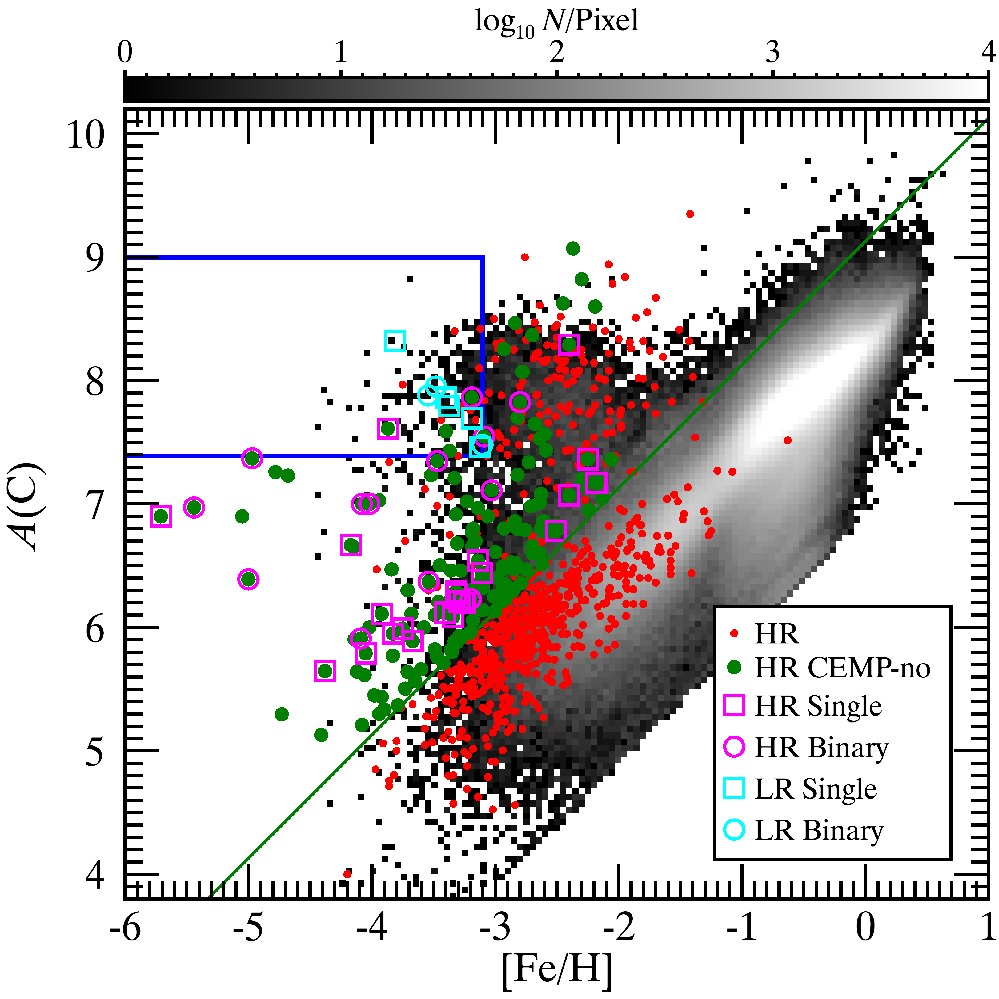}
\caption{\ac distribution of our LR sample, as a function of [Fe/H]. The black and white scale represents the logarithmic number density. The red dots are the stars from the HR sample; the green-filled circles are CEMP-no stars classified by [Ba/Fe] $<$ 0.0 from the HR sample. Note that stars with upper limits of [Ba/Fe] $>$ 0 are also regarded as CEMP-no stars. The green-solid line is at [C/Fe] = $+0.7$. The blue box is the region of interest for further consideration of our proposed group IV (see Section \ref{sec43} for the boundary definition). The magenta-open squares and circles are the recognized single and binary CEMP-no stars, respectively. The cyan-open squares and circles represent the recognized single and binary CEMP stars from our LR sample, respectively (see text).}
\label{fig4}
\end{figure}

In the LR sample (black points), we observe the expected increase of the overall CEMP fraction with decreasing metallicity: from $5\pm1$\% at [Fe/H] $\le$ --1.5 to $38\pm4$\% at [Fe/H] $\le$ --3.75, followed by a more rapid rise to $55^{+7}_{-8}$\% at [Fe/H] $\le$ --4.0. If the error is less than 1\%, we round it up to 1\%. These fractions are very similar to previous studies \citep[e.g.,][]{cohen2005,frebel2006,carollo2012,lee2013,placco2014,arentsen2023}. The HR sample (orange symbols) exhibits a similar trend but with a higher overall proportion and a steeper slope, increasing from $38\pm2$\% at [Fe/H] = --2.5 to $96^{+3}_{-6}$\% at [Fe/H] = --4.0.

When comparing the fractions of the CEMP subclasses, an interesting contrast emerges between the LR and HR samples. In the LR sample, both the \cemps\ (red squares) and CEMP-no (blue triangles) fractions increase with decreasing [Fe/H], with the \cemps\ fractions being slightly higher. In the HR sample, however, the CEMP-no fraction (magenta triangles) rises much more steeply than the \cemps\ fraction (green squares): the CEMP-no fraction increases from $24\pm2$\% at [Fe/H] = --2.5 to $83^{+7}_{-9}$\% at [Fe/H] = --4.0, while the \cemps\ fraction remains relatively flat at $\sim$11--13\%. This indicates a significantly higher CEMP-no fraction in the HR sample compared to the LR sample.

\citet{yoon2018} also presented CEMP-no and \cemps\ frequencies as a function of [Fe/H], using the AEGIS low-resolution spectroscopic sample, adopting \ac\ = 7.1 as the division. For [Fe/H] $<$ --2.0, they reported CEMP-no and \cemps\ fractions of 17\% and 10\%, respectively. The CEMP-no fraction increases to 35\%, 54\%, and 68\% for [Fe/H] $<$ --2.5, --3.0, and --3.5, respectively, while the \cemps\ fraction remains relatively stable at 6\%, 10\%, and 10\%. In our LR sample, the corresponding values for [Fe/H] $<$ --2.0 are $4\pm1$\% (CEMP-no) and $7\pm1$\% (\cemps). For [Fe/H] $<$ --2.5, --3.0, and --3.5, we obtain CEMP-no fractions of $10\pm1$\%, $16\pm1$\%, and $18\pm2$\%, respectively, showing no dramatic rise with decreasing metallicity. In contrast, our \cemps\ fractions are $14\pm1$\%, $19\pm1$\%, and $20\pm2$\% for these same metallicity cuts, higher than those reported by \citet{yoon2018}. Thus, there are clear differences between our LR sample compared with both the HR sample and the AEGIS sample of \citet{yoon2018}.

\begin{figure} [!t]
\begin{center}
\includegraphics[width=\columnwidth]{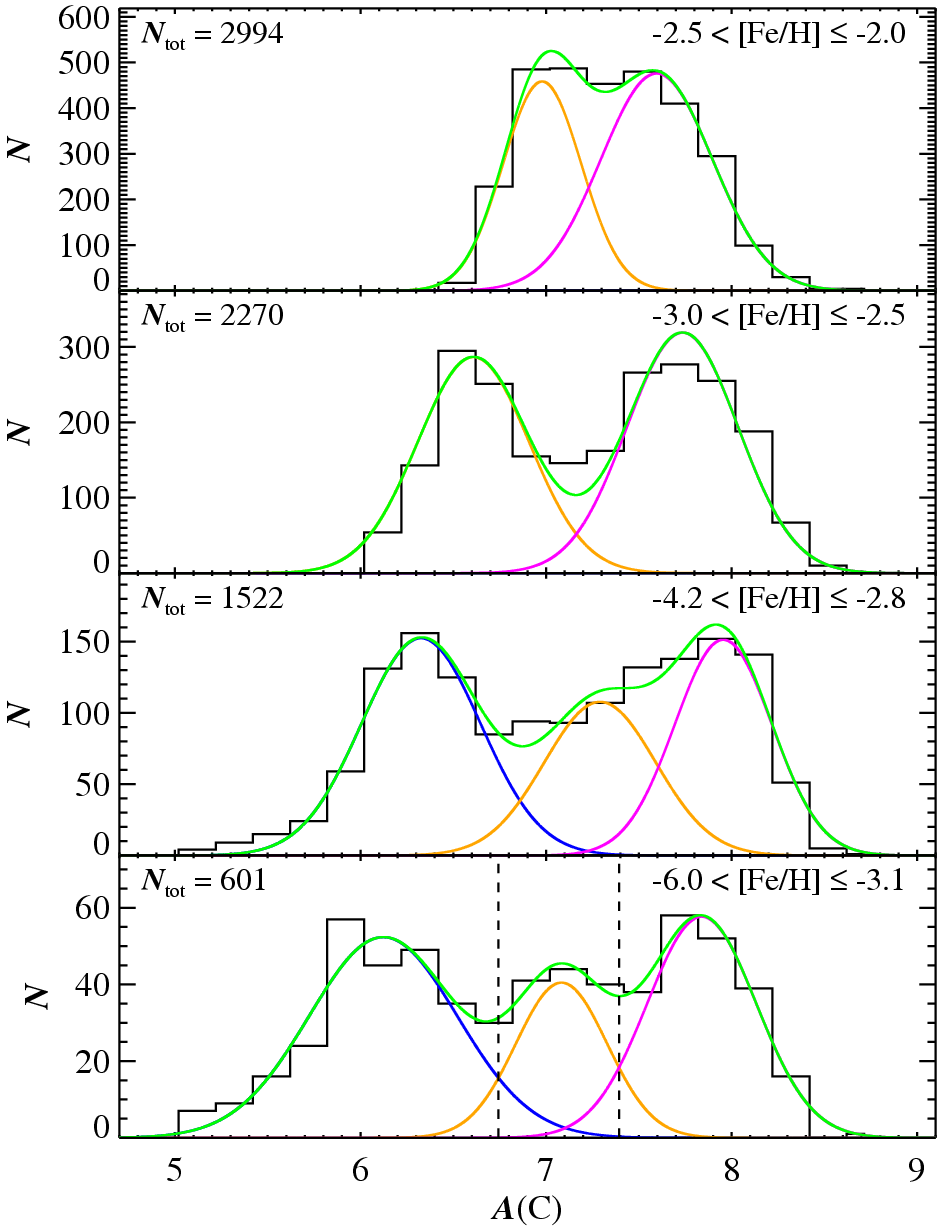}
\caption{\ac\ histograms of our LR sample in different ranges of [Fe/H]. Note that we adopt the HR measurements for overlapping stars between the LR and HR samples. The metallicity range is indicated at the top right of each panel. The bin size is 0.2 dex. $N_{\rm tot}$ is the total number of stars in each metallicity interval. The red, blue, and magenta curves are fits to the \ac histogram with a three-component Gaussian mixture model; the green line represents the sum of the three components. The separation of the three components becomes more evident as metallicity decreases. The dashed vertical lines shown in the lower panel are the divisions between the Group II and Group III stars (\ac = 6.74) and between the Group III and the proposed Group IV stars (\ac = 7.39), respectively, derived from the Gaussian mixture model.}
\label{fig5}
\end{center}
\end{figure}

\subsection{Evidence for a New High-\ac CEMP-no Group} \label{sec42}

Assuming the CEMP frequencies from the HR sample and \cite{yoon2018} are correct, how can we explain the contradictory pattern of the CEMP-no and CEMP-$s$ frequencies at low metallicity seen in Figure \ref{fig3}? We seek a resolution in the possible presence of other subclasses of CEMP-no stars.

It was first pointed out by \citet{rossi2005} that CEMP stars might not all share the same nucleosynthetic origin, based on the apparent bimodal distribution of [C/Fe] and $A$(C) at low metallicity (see their Figure 12). \citet{spite2013} and \citet{bonifacio2015} later argued that there existed high and low ``bands'' in plots of $A$(C) vs. [Fe/H] for main-sequence turnoff stars and mildly evolved subgiants; these bands were primarily associated with CEMP-$s$ and CEMP-no stars, respectively.

Using a significantly larger dataset of 305 CEMP stars from various literature sources based on high-resolution spectroscopic studies, \cite{yoon2016} demonstrated that the behavior of the CEMP stars in the $A$(C)--[Fe/H] space appeared to be more nuanced than could be captured by a simple two-band description. From Figure 1 of Yoon et al. -- the so-called Yoon-Beers Diagram -- these authors separated CEMP stars into three morphological groups: Group I with \cemps, CEMP-$r$, and CEMP-$r/s$ stars, Group II with CEMP-no stars with a clear correlation between \ac\ and [Fe/H], and Group III with CEMP-no stars without any apparent correlation between \ac and [Fe/H]. Based on this and the apparent differences between Group II and Group III stars in the $A$(Na)--$A$(C) and $A$(Mg)--$A$(C) spaces (Figure 4 of Yoon et al.), these authors argued for the existence of multiple progenitors and/or environments in which the CEMP-no stars formed.

\begin{deluxetable*}{lccc|ccccccc|cc}
\tablewidth{0in}
\renewcommand{\tabcolsep}{2pt}
\tablecaption{Group IV Stars from the HR Sample}
\tablehead{\colhead{} &\colhead{} &\colhead{} &\colhead{} & \multicolumn{7}{c}{High Resolution} & \multicolumn{2}{c}{Low Resolution} \\
\cline{5-11}  \cline{12-13} \\
\colhead{ID} & \colhead{R.A.} & \colhead{Dec.} & \colhead{$G$} & \colhead{\teff} & \colhead{\logg} & \colhead{[Fe/H]} & \colhead{[C/Fe]} & \colhead{$A$(C)} & \colhead{[Ba/Fe]} & \colhead{Binarity} & \colhead{[Fe/H]} & \colhead{[C/Fe]}  \\
\colhead{} & \colhead{(J2000)} & \colhead{(J2000)} & \colhead{(mag)} & \colhead{(K)} & \colhead{(cgs)} & \colhead{} & \colhead{} & \colhead{} & \colhead{} & \colhead{} & \colhead{}
} 
\startdata
CS~22957-027       &   23:59:13.14 &--03:53:48.4 & 13.39 &  5220 &  2.65 & --3.19 &  +2.62 &  7.86 & --0.81 &         2    & \nodata & \nodata  \\
CS~22958-042       &   02:01:07.44 &--57:16:58.9 & 14.40 &  5760 &  3.55 & --3.40 &  +2.56 &  7.59 & ~~--0.61$^{\rm a}$& 0 & \nodata & \nodata \\
CS~29498-043       &   21:03:52.10 &--29:42:50.2 & 13.35 &  4440 &  0.50 & --3.87 &  +3.05 &  7.61 & --0.49 &         1    &  \nodata & \nodata \\
HE~1456+0230       &   14:59:22.60 & +02:18:16.0 & 15.04 &  5664 &  2.20 & --3.37 &  +2.37 &  7.43 & --0.19 &         0    & --3.48  & +1.66   \\
\enddata
\tablecomments{Stellar parameters (\teff, \logg, and [Fe/H]), [C/Fe], and [Ba/Fe] for the High Resolution are adopted from \cite{yoon2016}. [C/Fe] and \ac values are corrected for evolutionary effects, following \citet{placco2014}. $G$ is the Gaia $G$ magnitude. In the Binarity column, 0 means ``unknown'', 1 for ``single'', and 2 for ``binary''. \\$^{\rm a}$ Upper limit of [Ba/Fe].}
\label{tab1}
\end{deluxetable*}

\begin{deluxetable}{crrrrrrr}
\tablewidth{0in}
\tabletypesize{\scriptsize}
\renewcommand{\tabcolsep}{2pt}
\tablecaption{Radial Velocity Variation of Group IV Candidates from the LR Sample}
\tablehead{\colhead{LAMOST} & \colhead{$RV_{\rm LR}$} & \colhead{$RV_{\rm G}$} & \colhead{$g_{\rm 0}$} & \colhead{\teff} & \colhead{\logg} & \colhead{[Fe/H]} & \colhead{[C/Fe]} \\
           \colhead{(J2000)      } & \colhead{(\kms)        } & \colhead{(\kms)        } & \colhead{ (mag) } & \colhead{(K)} & \colhead{(cgs)} & \colhead{} & \colhead{}} 
\startdata
J~040335.43--005510.0 &   --9.2$\pm$11.2 &  14.4$\pm$3.7 & 13.78 &  6286 &  3.73 & --3.12 & +2.16 \\
J~072314.49+073341.2  &    33.6$\pm$4.3 &   33.2$\pm$1.1 & 12.08 &  6374 &  3.04 & --3.37 & +2.74 \\
J~102749.53+290757.0  &   138.6$\pm$4.8 &  160.5$\pm$4.4 & 14.45 &  5274 &  1.60 & --3.81 & +3.70 \\
J~104538.77+580406.6  &  --56.6$\pm$3.1 &--125.0$\pm$2.2 & 13.52 &  5287 &  2.01 & --3.48 & +3.00 \\
J~152456.88+132655.9  &     9.2$\pm$4.0 & --87.5$\pm$12.6& 15.09 &  5146 &  2.02 & --3.54 & +3.00 \\
J~154135.19--005803.2 &  --38.0$\pm$4.5 & --32.3$\pm$4.1 & 15.29 &  5054 &  1.49 & --3.19 & +2.45 \\
J~155139.14+031426.3  &   --6.7$\pm$2.2 & --72.4$\pm$10.1& 13.12 &  4969 &  1.64 & --3.11 & +2.15 \\
J~174808.66+044600.7  & --168.6$\pm$4.2 &--158.2$\pm$7.5 & 13.47 &  5671 &  2.05 & --3.40 & +2.83 \\
\enddata
\tablecomments{[C/Fe] is corrected for evolutionary effects, following \citet{placco2014}. $RV_{\rm LR}$ is the heliocentric radial velocity from the LR sample, whereas $RV_{\rm G}$ is from $Gaia$.}
\label{tab2}
\end{deluxetable}

Following the logic of the Yoon et al. approach, we interpret the unexpected trends in the frequencies of CEMP-no and \cemps\ stars seen in Figure \ref{fig3} by positing a previously unrecognized subclass of CEMP-no stars. In particular, the unexpectedly higher fraction of \cemps\ stars below [Fe/H] = --3.0 may arise if high-\ac\ CEMP-no stars are classified as \cemps\ stars under our adopted definition. Figure \ref{fig4} shows our LR sample stars in the \ac\ vs. [Fe/H] diagram. In the figure, the black and white scale represents the logarithmic number density of our program stars. The red dots are the stars from the HR sample. The green-filled circles are the CEMP-no stars, based on the criterion of [Ba/Fe] $<$ 0\footnote[11]{We regard stars with an upper-limit estimate of [Ba/Fe] $>$ 0 as CEMP-no stars.}. The green-solid line is at [C/Fe] = +0.7. We also denote the recognized single stars among CEMP-no stars with low [Ba/Fe] as magenta-open squares, and recognized binary stars with low [Ba/Fe] as magenta-open circles.

In Figure \ref{fig4}, we note that some of the CEMP-no stars have \ac\ values larger than 7.0. Collectively, these stars were referred to as ``anomalous'' CEMP stars by \citet{yoon2016}. Based on the [Ba/Fe] abundance ratio, these objects are classified as CEMP-no stars. However, because of their high \ac\ value, some of them are categorized as \cemps\ in our definition (\ac\ $>$ 7.15). Consequently, if our sample includes many such stars, we might expect these to lead to the increasing frequency trend in the \cemps\ stars seen in Figure \ref{fig3}. Indeed, the figure indicates that our LR sample of CEMP stars includes many more high-\ac\ stars with \ac\ $>$ 7.0 for [Fe/H] $<$ --3.0 than those from the HR sample, indicating a possibility of a new group of CEMP stars.

We further investigate the possibility of a new group of CEMP stars in Figure \ref{fig5}, which presents \ac\ histograms of our LR sample across various metallicity bins, with particular emphasis on the distribution at [Fe/H] $\leq$ --3.1. The metallicity range for each panel is indicated at the top right of each plot. The blue, orange, and magenta lines represent components identified from a Gaussian mixture model, while the green line is their sum. 

It is evident that the CEMP stars in the high-metallicity region of Figure \ref{fig5} (top panel) are well-fit with two Gaussians. However, as the metallicity decreases, an additional group of CEMP stars is required to reproduce the \ac\ distribution. To be specific, from the sample of stars with [Fe/H] $\leq$ --3.1 (bottom panel), we derive peak values of \ac\ = 6.12 ($\sigma$ = 0.40 dex) for Group II (blue curve), \ac\ = 7.08 ($\sigma$ = 0.25 dex) for Group III (orange curve), and \ac\ = 7.84 ($\sigma$ = 0.29 dex) for a distinct high-\ac group (magenta curve). We also identify a dividing line (at the cross point) at \ac\ = 6.74 between Group II and Group III, and at \ac\ = 7.39 between Group III and the new high-\ac group. This clear separation strongly indicates the presence of a distinct population.

We can assess the statistical validity of this three-component interpretation for the \ac\ histogram using Gaussian mixture modeling for [Fe/H] $\leq$ --3.1. The three-component model yields the lowest BIC and AIC\footnote[12]{The Bayesian Information Criterion (BIC) and Akaike Information Criterion (AIC) are statistical measures used to compare competing models, balancing goodness of fit against model complexity to identify the least complex explanation of the data.}, outperforming the two-component model. This result provides strong statistical evidence for the existence of at least three distinct sub-populations among the CEMP stars at low metallicity: Group II, Group III, and a high-\ac\ group with a peak at \ac\ = 7.84. We refer to this high-\ac\ CEMP-no population as Group~IV, following the nomenclature of \cite{yoon2016}, and define it conservatively as stars with \ac\ $>$ 7.39, [Fe/H] $\leq$ --3.1, and [Ba/Fe] $<$ 0.0. Accordingly, Group III stars occupy the range \ac\ = 6.74--7.39.

The criterion of placing the division at \ac = 7.39, distinguishing between Group III and the proposed Group IV, is derived from the intersection of Gaussian components in the sample of stars with [Fe/H] $\leq$ --3.1. We can assess the robustness of this threshold with two tests. We first carry out a bootstrap resampling ($N$ = 10,000) of the \ac distribution, and find that the 1$\sigma$ variation in the intersection point is $\pm$0.08 dex, indicating statistical stability. Secondly, we verify that the trimodal structure remains and the intersection point is altered by less than 0.1 dex when using alternative bin sizes (0.15 or 0.25 dex). These checks indicate that the adopted boundary at \ac = 7.39 is unlikely to be an artifact of sample size, binning choice, or measurement uncertainty, and can be considered a robust criterion for separating Group IV stars from Group III stars.

\subsection{Properties of the New High-\ac CEMP-no Group} \label{sec43}

As established in Figure \ref{fig5}, the \ac\ distribution of CEMP stars at very low metallicity requires the presence of a third high-\ac\ component in addition to Groups II and III, which we designate as Group IV. We define this group by \ac\ $>$ 7.39, [Fe/H] $\leq$ --3.1, and [Ba/Fe] $<$ 0.0. Since [Ba/Fe] measurements are unavailable for the LR stars in the blue box of Figure \ref{fig4}, these objects are treated as Group IV candidates in this analysis. Note that we adopt the HR measurements for stars in common between the LR and HR samples. We now turn to quantifying the rate of occurrence of the Group IV stars and exploring their possible binary nature. These steps are essential for assessing the contribution of Group IV stars to the overall CEMP frequency trends and for constraining their likely formation channels.

We can estimate the fraction of Group IV stars using the HR sample. We focus on the blue box in Figure \ref{fig4}, defined by [Fe/H] $\leq$ --3.1 and \ac\ $>$ 7.39, which corresponds to the region of our adopted definition of Group IV. Within this region, there are 14 CEMP stars, of which four are CEMP-no stars (shown with green symbols), representing 29$^{+13}_{-10}$\% of the population. The stellar parameters, chemical abundances, and known binary status of these four stars are listed in Table \ref{tab1}. If we relax the selection to [Fe/H] $\leq$ --3.0 and \ac\ $\geq$ 7.26 -- 2$\sigma$ away from the Group IV peak derived from the bottom panel of Figure \ref{fig5} -- we find that 8 out of 26 stars are high-\ac\ CEMP-no stars, corresponding to 31$^{+10}_{-8}$\% of the total. Applying these fractions to our LR sample of stars in the blue box (black symbols) suggests that at least 30\% of them would be classified as Group IV.

\begin{figure} 
\centering
\epsscale{1.11}
\includegraphics[width=\columnwidth]{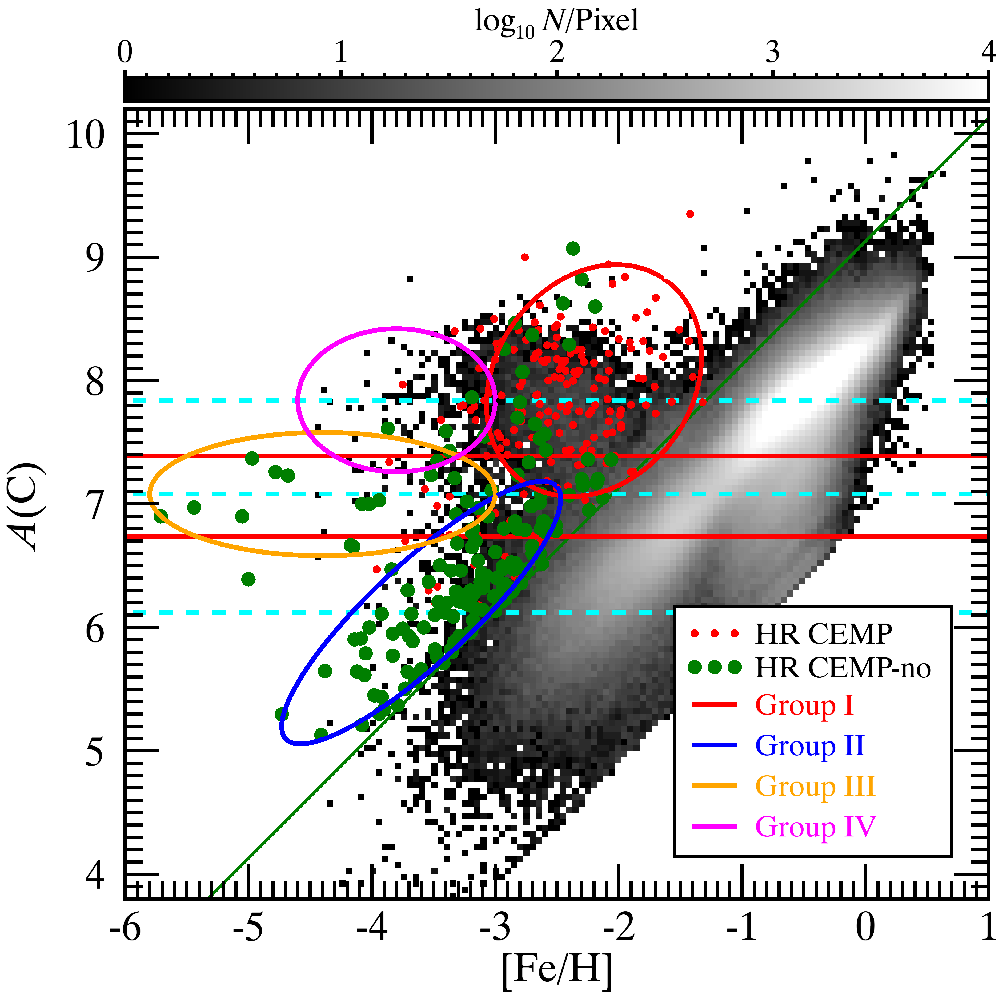}
\caption{The Yoon-Beers Diagram of the CEMP stars from our LR sample. Guided by the morphological scheme of \cite{yoon2016}, we identify Groups I, II, and III, as indicated by the colors of their ellipses. The magenta ellipse represents the region where the proposed new Group IV stars reside. The magnitudes of the minor axes of Group III and Group IV are set to 2$\sigma$ of the Gaussian fit to the \ac historam in the bottom panel of Figure \ref{fig5}. The cyan dashed lines indicate the peak of each group, while the red-solid lines are the dividing lines between the groups. Other symbols are the same as in Figure \ref{fig4}.}

\label{fig6}
\end{figure}

\begin{figure}
\centering
\includegraphics[width=\columnwidth]{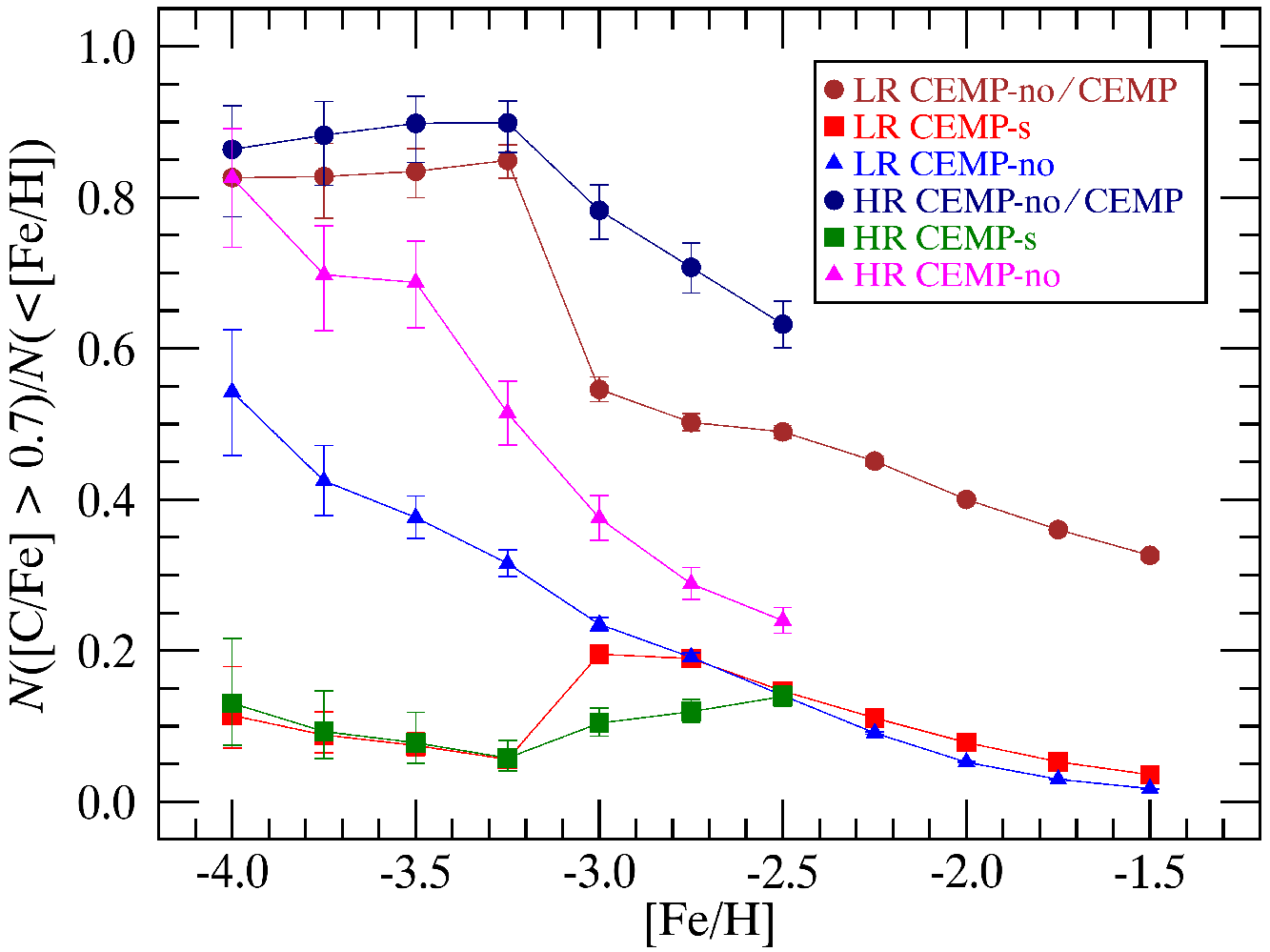}
\caption{Similar to Figure \ref{fig3}, but with Group IV candidates excluded (all stars in the blue box of Figure \ref{fig4} are treated as Group IV). Note that brown (LR CEMP-no/CEMP) and navy (HR CEMP-no/CEMP) symbols indicate the cumulative fractions of CEMP-no stars relative to \textit{all} CEMP stars for the LR and HR samples, respectively (see text for details).}
\label{fig7}
\end{figure}

To investigate the binary nature of these stars, we cross-match the LR CEMP stars located in the blue box of Figure \ref{fig4} with \gaia\ DR3 \citep{gaia2023} to obtain RV measurements. Eight matches are found, and their RV data are listed in Table \ref{tab2}. Adopting a conservative criterion for binary identification, we identify three stars (38$^{+18}_{-15}$\%) that exhibit RV offsets larger than 30 \kms\ between our LR measurements and the \gaia\ values, indicating a possible binary nature. These stars are shown as cyan-open circles in Figure \ref{fig4}, while cyan-open squares mark candidates with RV differences smaller than 30 \kms.

In the HR sample, one confirmed binary (magenta-open square) is identified among the four CEMP-no stars in the blue box of Figure~\ref{fig4}, corresponding to a binary fraction of 25$^{+25}_{-15}$\%. This fraction is lower than that of the LR sample, where three out of eight stars are binaries. As noted by \citet{aguado2023}, additional binaries may still be present among CEMP-no stars, particularly wide systems whose RV variations are difficult to detect without long-term monitoring. Assuming all eight LR stars are CEMP-no based on their [Ba/Fe] abundances, the combined HR and LR samples yield an overall binary fraction of 33$^{+14}_{-12}$\% for Group~IV stars.

Figure \ref{fig6} illustrates the Yoon-Beers diagram, where colored ellipses mark the areas corresponding to Groups I, II, and III according to the classification system of \citet{yoon2016}. The magenta ellipse identifies the location of Group IV stars. CEMP-no stars (depicted as green-filled circles) and \cemps\ stars (represented by red dots) from the HR sample -- classified using the [Ba/Fe] criterion -- are also shown in the figure for context. The cyan-dashed lines indicate the peak positions of each group, while the red-solid lines define the separating boundaries. A notable difference is observed: Group III stars reach much lower metallicities, whereas Group IV candidates are found down to [Fe/H] $\approx$ --4.5 at most. Similar to Group III, Group IV candidates exhibit no apparent correlation between \ac\ and [Fe/H], which might provide insight into their progenitor systems.

Figure \ref{fig7} shows the cumulative frequencies of CEMP-no and \cemps\ stars after excluding Group IV candidates (all stars in the blue box of Figure \ref{fig4}). Blue and red symbols correspond to the LR sample, while magenta and green symbols represent the HR sample. Once Group IV candidates are removed, the \cemps\ fractions in both datasets converge to 5--13\% with very similar trends for [Fe/H] $< -3.1$, in good agreement with the values reported by \citet{yoon2018}.

For the CEMP-no fraction, the LR sample (blue symbols) increases from 25$^{+7}_{-6}$\% (from Figure \ref{fig3}) to 31$^{+8}_{-7}$\% at [Fe/H] = $-4.0$, but remains systematically below the HR and AEGIS results of \cite{yoon2018} (over 80\%). This discrepancy is mainly due to sample selection and detection thresholds. As mentioned in Section \ref{sec41}, the LR CEMP fractions represent lower limits because only CEMP stars with strong detections (CC $>$ 0.5) are classified as CEMP, whereas those with CC $<$ 0.5 are counted as non-CEMP. In addition, the LR sample includes many dwarfs, while subgiants and giants dominate the AEGIS and HR samples.

To mitigate such biases, we compare the cumulative fractions of CEMP-no stars relative to all CEMP stars (brown symbols for LR and navy for HR in Figure \ref{fig7}). Although the HR fractions remain higher by $\sim$10\%, both datasets follow similar rising trends with decreasing metallicity. Relaxing the detection threshold of the CEMP stars to CC $>$ 0.3 further improves the agreement: the LR CEMP-no frequency reaches 54$\pm8$\% at [Fe/H] $< -4.0$, and the CEMP-no fractions over all CEMP stars are 83--85\% (errors 2--9\%) for LR and 86--90\% (errors 3--9\%) for HR at [Fe/H] $< -3.1$. These results demonstrate that the apparent offsets between LR and HR samples arise from selection effects, while the underlying trends are consistent within the uncertainties. The lower CEMP-no fraction of the LR sample (Figure \ref{fig3}) can be attributed to both sample selection biases and inclusion of Group IV candidates, which are misclassified as \cemps\ when using the \ac-based criterion.

\section{Discussion and Conclusions}\label{sec5}

Our analysis of the HR sample indicates that nearly one-third of stars with high \ac\ ($>$ 7.39) and very low metallicity ([Fe/H] $\leq$ --3.1) belong to the CEMP-no subclass rather than CEMP-$s$. Extrapolating this fraction to the LR dataset suggests that a comparable proportion of stars within the blue box in Figure~\ref{fig4} can also be classified as high-\ac\ CEMP-no stars, which we refer to as Group IV. Notably, the combined HR and LR data (Group IV candidates) point to a binary fraction of 33$^{+14}_{-12}$\% for this group. These findings motivate a closer examination of the possible progenitors of Group IV stars.

The presence of multiple subclasses among CEMP stars implies that a variety of formation pathways are required to explain their diverse chemical-abundance patterns. It is generally accepted that CEMP-no and CEMP-$s$ stars arise from distinct astrophysical progenitors. Specifically, CEMP-no stars are thought to originate from: (1) “faint SNe” or “mixing-and-fallback” explosions of $\sim$20--60~\msun\ progenitors \citep{umeda2003,umeda2005,nomoto2013,tominaga2014}, or (2) rapidly rotating, massive, metal-free stars with masses $>$ 60--100~\msun\ (“spinstars”; \citealt{meynet2006,meynet2010,chiappini2013}). By contrast, CEMP-$s$ stars are believed to form via mass transfer in binary systems, where an intermediate-mass (1--4~\msun) AGB star enriches its companion with $s$-process elements \citep{suda2004,herwig2005,lucatello2005,komiya2007,bisterzo2011,hansen2015}. Long-term RV-monitoring studies \citep{starkenburg2014,hansen2016a,hansen2016b,jorissen2016} have revealed markedly different binary fractions for these two subclasses, providing strong observational support for these distinct origins.

Within the CEMP-no category, the origins of the Group II and Group III stars defined by \cite{yoon2016} remain debated. Their abundance patterns can be reproduced either by faint SNe \citep[e.g.,][]{iwamoto2005,ishigaki2014,salvadori2015,komiya2020} or by spinstars \citep[e.g.,][]{meynet2006,liu2021}. Meanwhile, other models avoid invoking Population III (Pop III) progenitors altogether, suggesting that CEMP-no stars could form from gas enriched by Pop II AGB stars \citep[e.g.,][]{sharma2019} or by normal SNe II \citep{jeon2021}. \cite{chiaki2017} further argued that differences in dominant gas coolants during star formation could produce the observed subclasses. The drivers of these sub-populations thus remain uncertain.

Based on the observed dependence of \ac\ on [Fe/H], \cite{yoon2016} suggested that Group II stars may originate from mixing-and-fallback faint SNe, whereas Group III stars may be associated with spinstars. However, the high \ac\ values of the Group~IV stars identified here cannot be easily explained by either model, indicating the need for additional or hybrid enrichment mechanisms.

At first glance, Groups III and IV might appear to represent a continuous sequence. Yet, Group III stars extend to much lower metallicities than Group IV candidates (see Figure\ref{fig4}), and the \ac\ distribution of Group IV exhibits statistically significant peaks, both pointing to a distinct origin. In the blue box of Figure \ref{fig4}, we also identify one binary CEMP-no star (magenta-open circle) from the HR sample and three binary candidates (cyan-open circles) from RV differences between the LR and \gaia\ data, corresponding to a binary fraction of roughly 30\%. Although the sample is small, these results support the interpretation of Group IV as a distinct subclass of CEMP-no stars, with at least some members potentially arising from Pop II AGB mass transfer that yields little $n$-capture material in a low-metallicity environment \citep[e.g.,][]{suda2004,lau2007}. Additional channels, such as Pop~II AGB winds or incomplete AGB mass transfer, may also contribute.

The $^{12}$C/$^{13}$C ratio is often used as an indicator of AGB mass transfer, as this ratio tends to decrease due to CN cycling during the red giant branch phase. In this context, the incomplete AGB mass transfer scenario is particularly relevant for the Group~IV binary CS~22957-027, which shows a measured $^{12}$C/$^{13}$C ratio of 8$\pm$2 \citep{norris1997,bonifacio1998,aoki2002}. Such a low ratio is consistent with partial AGB mass transfer, where CN-processed material from the donor reduces $^{12}$C/$^{13}$C while enhancing \ac\ \citep[e.g.,][]{lucatello2003,stancliffe2009}. The absence of $n$-capture enrichment, however, points to non-standard or atypical transfer pathways.

As shown in Table~\ref{tab1}, CS~22958-042 and CS~29498-043 also exhibit low $^{12}$C/$^{13}$C ratios ($<$ 10) \citep{aoki2002,sivarani2006,molaro2023}. CS~29498-043 appears to be single, while the binary status of CS~22958-42 remains uncertain but may also be single. For such stars, the low isotopic ratios are more plausibly explained by natal enrichment from CN-processed material ejected by faint SNe with fallback \citep{umeda2003} or by rapidly rotating massive stars (spinstars; \citealt{meynet2006}), even though spinstars are expected to produce more $^{13}$C than metal-free faint SNe \citep{meynet2006,heger2010,limongi2018}. This overlap in isotopic signatures demonstrates that low $^{12}$C/$^{13}$C is not restricted to binaries, and underscores the need to combine isotopic ratios with additional abundance indicators (e.g., [Li/Fe], [N/Fe], [O/Fe], [Na/Fe]) as well as long-term RV monitoring to disentangle their formation pathways. Nevertheless, the possibility remains that some of these stars (or Group IV stars) are very wide binaries whose RV variations are too small to detect without extended monitoring \citep[e.g.,][]{aguado2023}. In such cases, the absence of $n$-capture enrichment in their atmospheres is again particularly difficult to reconcile.

Previous work reinforces the above view. \citet{hansen2015} identified several high-\ac\ CEMP-no stars (binary status unknown), highlighting that \ac\ bimodality cannot be explained solely by the binary vs. single-star dichotomy. More recently, \citet{arentsen2019} found a higher binary fraction among high-\ac\ CEMP-no stars (47$^{+15}_{-14}$\%) than among low-\ac\ stars (18$^{+14}_{-9}$\%), a trend broadly consistent with our estimate of 33$^{+14}_{-12}$\% for Group IV. Together, these results suggest that Group IV stars represent a distinct subclass of CEMP-no stars whose origins likely involve a mixture of formation pathways -- ranging from extrinsic AGB transfer to intrinsic natal enrichment by faint SNe or spinstars.

\cite{hartwig2018} proposed a framework for identifying mono-enriched second-generation stars formed from a single Pop III SN. Groups II and III are broadly consistent with this picture, but the inclusion of binaries within Group~IV suggests that some stars have undergone atypical AGB mass transfer without $s$-process enrichment. This points to hybrid or multiple enrichment pathways in the early Universe.

Finally, \cite{rossi2023,rossi2024} modeled the Bo\"{o}tes I dwarf galaxy, including all carbon sources from SNe and AGB stars of Pop III and Pop II progenitors. Their results indicate that stars with \ac\ $>$ 6.0 are primarily enriched by Pop II AGB stars, while true Pop III descendants have \ac\ $<$ 6.0. They also proposed a subclass of moderate \cemps\ stars (\ac\ $\sim$ 7.0, 0 $<$ [Ba/Fe] $<$ 1) from Pop II AGB winds without mass transfer. The \ac\ range predicted for Pop II descendants in \citet{rossi2024} overlaps with that of our Group IV candidates, suggesting that some may belong to this category.

The progenitors of Group IV CEMP-no stars remain uncertain. Binary systems may reflect inefficient mass transfer from low-metallicity AGB companions, while the origin of single stars with the very high \ac\ ($>$ 7.84, the Group IV peak) is especially difficult to explain with standard scenarios such as faint Pop III SNe or spinstars. These stars likely require alternative or hybrid enrichment channels, including atypical Pop II/III progenitors, incomplete ejecta mixing, or fallback-limited explosions. A promising explanation is early contributions from intermediate-mass AGB stars, whose nucleosynthesis can produce strong C and N enhancements with only weak or diluted $s$-process signatures, as recently suggested for the hyper metal-poor star HE 1327-2326 \citep{gilpons2025}. Although that star is recognized as a \cemps\ object with enhanced Ba, its case shows that AGB stars may already have influenced the earliest chemical evolution. Alternatively, Group IV may reflect inhomogeneous mixing of ejecta from multiple sources -- AGB stars, faint SNe, and rotating massive stars -- in the early ISM. High-resolution measurements of key $n$-capture elements such as Sr, Ba, and Eu will be critical to distinguish between these possibilities.

Our results provide strong evidence for a new subclass of CEMP-no stars with unusually high \ac\ at extremely low metallicity, but several limitations remain. Incomplete knowledge of binarity limits assessment of mass-transfer contributions, and classification based on low-resolution spectra and \ac\ alone is uncertain without [Ba/Fe] constraints. In addition, differences in subclass frequencies compared to previous work may reflect varying selection criteria or methods. Future high-resolution spectroscopy, long-term RV monitoring, and improved theoretical models will be essential to clarify the origins and broader significance of this newly identified stellar population.


We thank an anonymous referee for a careful review of this paper, which has significantly improved the clarity of presentation. Y.S.L. acknowledges support from the National Research Foundation (NRF) of Korea grant funded by the Ministry of Science and ICT (RS-2024-00333766). Y.S.L. also thanks J. Yoon for reviewing this manuscript and suggesting helpful comments to improve this paper. T.C.B. and J.H. acknowledge partial support for this work from grant PHY 14-30152; Physics Frontier Center/JINA Center for the Evolution of the Elements (JINA-CEE), and OISE-1927130: The International Research Network for Nuclear Astrophysics (IReNA), awarded by the US National Science Foundation. Y.H. acknowledges support from the JSPS KAKENHI Grant Numbers JP22KJ0157, JP25H00664, and JP25K01046.

Funding for the Sloan Digital Sky Survey IV has been provided by the Alfred P. Sloan Foundation, the U.S. Department of Energy Office of Science, and the Participating Institutions.

This work presents results from the European Space Agency (ESA) space mission Gaia. Gaia data are being processed by the Gaia Data Processing and Analysis Consortium (DPAC). Funding for the DPAC is provided by national institutions, in particular the institutions participating in the Gaia MultiLateral Agreement (MLA). The Gaia mission website is \url{https://www.cosmos.esa.int/gaia}. The Gaia archive website is \url{https://archives.esac.esa.int/gaia.}

The Guoshoujing Telescope (the Large Sky Area Multi-Object Fiber Spectroscopic Telescope, LAMOST) is a National Major Scientific Project which is built by the Chinese Academy of Sciences, funded by the National Development and Reform Commission, and operated and managed by the National Astronomical Observatories, Chinese Academy of Sciences.


\vfill\eject

\bibliographystyle{aasjournal}
\bibliography{ref.bib}{}

\begin{thebibliography}{}
\expandafter\ifx\csname natexlab\endcsname\relax\def\natexlab#1{#1}\fi
\providecommand{\url}[1]{\href{#1}{#1}}
\providecommand{\dodoi}[1]{doi:~\href{http://doi.org/#1}{\nolinkurl{#1}}}
\providecommand{\doeprint}[1]{\href{http://ascl.net/#1}{\nolinkurl{http://ascl.net/#1}}}
\providecommand{\doarXiv}[1]{\href{https://arxiv.org/abs/#1}{\nolinkurl{https://arxiv.org/abs/#1}}}

\bibitem[{{Aguado} {et~al.}(2023){Aguado}, {Caffau}, {Molaro}, {Allende Prieto}, {Bonifacio}, {Gonz{\'a}lez Hern{\'a}ndez}, {Rebolo}, {Salvadori}, {Zapatero Osorio}, {Cristiani}, {Pepe}, {Santos}, {Cupani}, {Di Marcantonio}, {D'Odorico}, {Lovis}, {Nunes}, {Martins}, {Milakovi}, {Rodrigues}, {Schmidt}, {Sozzetti}, \& {Su{\'a}rez Mascare{\~n}o}}]{aguado2023}
{Aguado}, D.~S., {Caffau}, E., {Molaro}, P., {et~al.} 2023, \aap, 669, L4, \dodoi{10.1051/0004-6361/202245392}

\bibitem[{{Allende Prieto} {et~al.}(2008){Allende Prieto}, {Sivarani}, {Beers}, {Lee}, {Koesterke}, {Shetrone}, {Sneden}, {Lambert}, {Wilhelm}, {Rockosi}, {Lai}, {Yanny}, {Ivans}, {Johnson}, {Aoki}, {Bailer-Jones}, \& {Re Fiorentin}}]{allende2008}
{Allende Prieto}, C., {Sivarani}, T., {Beers}, T.~C., {et~al.} 2008, \aj, 136, 2070, \dodoi{10.1088/0004-6256/136/5/2070}

\bibitem[{{Aoki} {et~al.}(2007){Aoki}, {Beers}, {Christlieb}, {Norris}, {Ryan}, \& {Tsangarides}}]{aoki2007}
{Aoki}, W., {Beers}, T.~C., {Christlieb}, N., {et~al.} 2007, \apj, 655, 492, \dodoi{10.1086/509817}

\bibitem[{{Aoki} {et~al.}(2002){Aoki}, {Norris}, {Ryan}, {Beers}, \& {Ando}}]{aoki2002}
{Aoki}, W., {Norris}, J.~E., {Ryan}, S.~G., {Beers}, T.~C., \& {Ando}, H. 2002, \pasj, 54, 933, \dodoi{10.1093/pasj/54.6.933}

\bibitem[{{Ardern-Arentsen} {et~al.}(2025){Ardern-Arentsen}, {Kane}, {Belokurov}, {Matsuno}, {Montelius}, {Monty}, \& {Sanders}}]{arentsen2025}
{Ardern-Arentsen}, A., {Kane}, S.~G., {Belokurov}, V., {et~al.} 2025, \mnras, 537, 1984, \dodoi{10.1093/mnras/staf096}

\bibitem[{{Arentsen} {et~al.}(2023){Arentsen}, {Aguado}, {Sestito}, {Gonz{\'a}lez Hern{\'a}ndez}, {Martin}, {Starkenburg}, {Jablonka}, \& {Yuan}}]{arentsen2023}
{Arentsen}, A., {Aguado}, D.~S., {Sestito}, F., {et~al.} 2023, \mnras, 519, 5554, \dodoi{10.1093/mnras/stad043}

\bibitem[{{Arentsen} {et~al.}(2022){Arentsen}, {Placco}, {Lee}, {Aguado}, {Martin}, {Starkenburg}, \& {Yoon}}]{arentsen2022}
{Arentsen}, A., {Placco}, V.~M., {Lee}, Y.~S., {et~al.} 2022, \mnras, 515, 4082, \dodoi{10.1093/mnras/stac2062}

\bibitem[{{Arentsen} {et~al.}(2019){Arentsen}, {Starkenburg}, {Shetrone}, {Venn}, {Depagne}, \& {McConnachie}}]{arentsen2019}
{Arentsen}, A., {Starkenburg}, E., {Shetrone}, M.~D., {et~al.} 2019, \aap, 621, A108, \dodoi{10.1051/0004-6361/201834146}

\bibitem[{{Arentsen} {et~al.}(2021){Arentsen}, {Starkenburg}, {Aguado}, {Martin}, {Placco}, {Carlberg}, {Gonz{\'a}lez Hern{\'a}ndez}, {Hill}, {Jablonka}, {Kordopatis}, {Lardo}, {Mashonkina}, {Navarro}, {Venn}, {Buder}, {Lewis}, {Wan}, \& {Zucker}}]{arentsen2021}
{Arentsen}, A., {Starkenburg}, E., {Aguado}, D.~S., {et~al.} 2021, \mnras, 505, 1239, \dodoi{10.1093/mnras/stab1343}

\bibitem[{{Beers} \& {Christlieb}(2005)}]{beers2005}
{Beers}, T.~C., \& {Christlieb}, N. 2005, \araa, 43, 531, \dodoi{10.1146/annurev.astro.42.053102.134057}

\bibitem[{{Bisterzo} {et~al.}(2011){Bisterzo}, {Gallino}, {Straniero}, {Cristallo}, \& {K{\"a}ppeler}}]{bisterzo2011}
{Bisterzo}, S., {Gallino}, R., {Straniero}, O., {Cristallo}, S., \& {K{\"a}ppeler}, F. 2011, \mnras, 418, 284, \dodoi{10.1111/j.1365-2966.2011.19484.x}

\bibitem[{{Bisterzo} {et~al.}(2012){Bisterzo}, {Gallino}, {Straniero}, {Cristallo}, \& {K{\"a}ppeler}}]{bisterzo2012}
---. 2012, \mnras, 422, 849, \dodoi{10.1111/j.1365-2966.2012.20670.x}

\bibitem[{{Blanton} {et~al.}(2017){Blanton}, {Bershady}, {Abolfathi}, {Albareti}, {Allende Prieto}, {Almeida}, {Alonso-Garc{\'\i}a}, {Anders}, {Anderson}, {Andrews}, {Aquino-Ort{\'\i}z}, {Arag{\'o}n-Salamanca}, {Argudo-Fern{\'a}ndez}, {Armengaud}, {Aubourg}, {Avila-Reese}, {Badenes}, {Bailey}, {Barger}, {Barrera-Ballesteros}, {Bartosz}, {Bates}, {Baumgarten}, {Bautista}, {Beaton}, {Beers}, {Belfiore}, {Bender}, {Berlind}, {Bernardi}, {Beutler}, {Bird}, {Bizyaev}, {Blanc}, {Blomqvist}, {Bolton}, {Boquien}, {Borissova}, {van den Bosch}, {Bovy}, {Brandt}, {Brinkmann}, {Brownstein}, {Bundy}, {Burgasser}, {Burtin}, {Busca}, {Cappellari}, {Delgado Carigi}, {Carlberg}, {Carnero Rosell}, {Carrera}, {Chanover}, {Cherinka}, {Cheung}, {G{\'o}mez Maqueo Chew}, {Chiappini}, {Choi}, {Chojnowski}, {Chuang}, {Chung}, {Cirolini}, {Clerc}, {Cohen}, {Comparat}, {da Costa}, {Cousinou}, {Covey}, {Crane}, {Croft}, {Cruz-Gonzalez}, {Garrido Cuadra}, {Cunha}, {Damke}, {Darling}, {Davies}, {Dawson}, {de la Macorra}, {Dell'Agli}, {De
  Lee}, {Delubac}, {Di Mille}, {Diamond-Stanic}, {Cano-D{\'\i}az}, {Donor}, {Downes}, {Drory}, {du Mas des Bourboux}, {Duckworth}, {Dwelly}, {Dyer}, {Ebelke}, {Eigenbrot}, {Eisenstein}, {Emsellem}, {Eracleous}, {Escoffier}, {Evans}, {Fan}, {Fern{\'a}ndez-Alvar}, {Fernandez-Trincado}, {Feuillet}, {Finoguenov}, {Fleming}, {Font-Ribera}, {Fredrickson}, {Freischlad}, {Frinchaboy}, {Fuentes}, {Galbany}, {Garcia-Dias}, {Garc{\'\i}a-Hern{\'a}ndez}, {Gaulme}, {Geisler}, {Gelfand}, {Gil-Mar{\'\i}n}, {Gillespie}, {Goddard}, {Gonzalez-Perez}, {Grabowski}, {Green}, {Grier}, {Gunn}, {Guo}, {Guy}, {Hagen}, {Hahn}, {Hall}, {Harding}, {Hasselquist}, {Hawley}, {Hearty}, {Gonzalez Hern{\'a}ndez}, {Ho}, {Hogg}, {Holley-Bockelmann}, {Holtzman}, {Holzer}, {Huehnerhoff}, {Hutchinson}, {Hwang}, {Ibarra-Medel}, {da Silva Ilha}, {Ivans}, {Ivory}, {Jackson}, {Jensen}, {Johnson}, {Jones}, {J{\"o}nsson}, {Jullo}, {Kamble}, {Kinemuchi}, {Kirkby}, {Kitaura}, {Klaene}, {Knapp}, {Kneib}, {Kollmeier}, {Lacerna}, {Lane}, {Lang}, {Law},
  {Lazarz}, {Lee}, {Le Goff}, {Liang}, {Li}, {Li}, {Lian}, {Lima}, {Lin}, {Lin}, {Bertran de Lis}, {Liu}, {de Icaza Lizaola}, {Long}, {Lucatello}, {Lundgren}, {MacDonald}, {Deconto Machado}, {MacLeod}, {Mahadevan}, {Geimba Maia}, {Maiolino}, {Majewski}, {Malanushenko}, {Malanushenko}, {Manchado}, {Mao}, {Maraston}, {Marques-Chaves}, {Masseron}, {Masters}, {McBride}, {McDermid}, {McGrath}, {McGreer}, {Medina Pe{\~n}a}, \& {Melendez}}]{blanton2017}
{Blanton}, M.~R., {Bershady}, M.~A., {Abolfathi}, B., {et~al.} 2017, \aj, 154, 28, \dodoi{10.3847/1538-3881/aa7567}

\bibitem[{{Bonifacio} {et~al.}(1998){Bonifacio}, {Molaro}, {Beers}, \& {Vladilo}}]{bonifacio1998}
{Bonifacio}, P., {Molaro}, P., {Beers}, T.~C., \& {Vladilo}, G. 1998, \aap, 332, 672, \dodoi{10.48550/arXiv.astro-ph/9712227}

\bibitem[{{Bonifacio} {et~al.}(2015){Bonifacio}, {Caffau}, {Spite}, {Limongi}, {Chieffi}, {Klessen}, {Fran{\c{c}}ois}, {Molaro}, {Ludwig}, {Zaggia}, {Spite}, {Plez}, {Cayrel}, {Christlieb}, {Clark}, {Glover}, {Hammer}, {Koch}, {Monaco}, {Sbordone}, \& {Steffen}}]{bonifacio2015}
{Bonifacio}, P., {Caffau}, E., {Spite}, M., {et~al.} 2015, \aap, 579, A28, \dodoi{10.1051/0004-6361/201425266}

\bibitem[{{Carney} {et~al.}(2003){Carney}, {Latham}, {Stefanik}, {Laird}, \& {Morse}}]{carney2003}
{Carney}, B.~W., {Latham}, D.~W., {Stefanik}, R.~P., {Laird}, J.~B., \& {Morse}, J.~A. 2003, \aj, 125, 293, \dodoi{10.1086/345386}

\bibitem[{{Carollo} {et~al.}(2014){Carollo}, {Freeman}, {Beers}, {Placco}, {Tumlinson}, \& {Martell}}]{carollo2014}
{Carollo}, D., {Freeman}, K., {Beers}, T.~C., {et~al.} 2014, \apj, 788, 180, \dodoi{10.1088/0004-637X/788/2/180}

\bibitem[{{Carollo} {et~al.}(2012){Carollo}, {Beers}, {Bovy}, {Sivarani}, {Norris}, {Freeman}, {Aoki}, {Lee}, \& {Kennedy}}]{carollo2012}
{Carollo}, D., {Beers}, T.~C., {Bovy}, J., {et~al.} 2012, \apj, 744, 195, \dodoi{10.1088/0004-637X/744/2/195}

\bibitem[{{Chiaki} {et~al.}(2017){Chiaki}, {Tominaga}, \& {Nozawa}}]{chiaki2017}
{Chiaki}, G., {Tominaga}, N., \& {Nozawa}, T. 2017, \mnras, 472, L115, \dodoi{10.1093/mnrasl/slx163}

\bibitem[{{Chiappini}(2013)}]{chiappini2013}
{Chiappini}, C. 2013, Astronomische Nachrichten, 334, 595, \dodoi{10.1002/asna.201311902}

\bibitem[{{Chiti} {et~al.}(2018){Chiti}, {Simon}, {Frebel}, {Thompson}, {Shectman}, {Mateo}, {Bailey}, {Crane}, \& {Walker}}]{chiti2018}
{Chiti}, A., {Simon}, J.~D., {Frebel}, A., {et~al.} 2018, \apj, 856, 142, \dodoi{10.3847/1538-4357/aab663}

\bibitem[{{Choplin} {et~al.}(2017){Choplin}, {Ekstr{\"o}m}, {Meynet}, {Maeder}, {Georgy}, \& {Hirschi}}]{choplin2017}
{Choplin}, A., {Ekstr{\"o}m}, S., {Meynet}, G., {et~al.} 2017, \aap, 605, A63, \dodoi{10.1051/0004-6361/201630182}

\bibitem[{{Choplin} {et~al.}(2016){Choplin}, {Maeder}, {Meynet}, \& {Chiappini}}]{choplin2016}
{Choplin}, A., {Maeder}, A., {Meynet}, G., \& {Chiappini}, C. 2016, \aap, 593, A36, \dodoi{10.1051/0004-6361/201628083}

\bibitem[{{Choplin} {et~al.}(2024){Choplin}, {Siess}, {Goriely}, \& {Martinet}}]{choplin2024}
{Choplin}, A., {Siess}, L., {Goriely}, S., \& {Martinet}, S. 2024, \aap, 684, A206, \dodoi{10.1051/0004-6361/202348957}

\bibitem[{{Cohen} {et~al.}(2005){Cohen}, {Shectman}, {Thompson}, {McWilliam}, {Christlieb}, {Melendez}, {Zickgraf}, {Ram{\'\i}rez}, \& {Swenson}}]{cohen2005}
{Cohen}, J.~G., {Shectman}, S., {Thompson}, I., {et~al.} 2005, \apjl, 633, L109, \dodoi{10.1086/498502}

\bibitem[{{Cowan} \& {Rose}(1977)}]{cowan1977}
{Cowan}, J.~J., \& {Rose}, W.~K. 1977, ApJ, 212, 149, \dodoi{10.1086/155030}

\bibitem[{{Cui} {et~al.}(2012){Cui}, {Zhao}, {Chu}, {Li}, {Li}, {Zhang}, {Su}, {Yao}, {Wang}, {Xing}, {Li}, {Zhu}, {Wang}, {Gu}, {Luo}, {Xu}, {Zhang}, {Liu}, {Zhang}, {Yang}, {Cao}, {Chen}, {Chen}, {Chen}, {Chen}, {Chu}, {Feng}, {Gong}, {Hou}, {Hu}, {Hu}, {Hu}, {Jia}, {Jiang}, {Jiang}, {Jiang}, {Jin}, {Li}, {Li}, {Li}, {Liu}, {Liu}, {Lu}, {Mao}, {Men}, {Qi}, {Qi}, {Shi}, {Tang}, {Tao}, {Wang}, {Wang}, {Wang}, {Wang}, {Wang}, {Wang}, {Wang}, {Wang}, {Wang}, {Wang}, {Wang}, {Wang}, {Xu}, {Xu}, {Yang}, {Yu}, {Yuan}, {Yuan}, {Zhai}, {Zhang}, {Zhang}, {Zhang}, {Zhao}, {Zhou}, {Zhou}, {Zhu}, \& {Zou}}]{cui2012}
{Cui}, X.-Q., {Zhao}, Y.-H., {Chu}, Y.-Q., {et~al.} 2012, Research in Astronomy and Astrophysics, 12, 1197, \dodoi{10.1088/1674-4527/12/9/003}

\bibitem[{{Dawson} {et~al.}(2013){Dawson}, {Schlegel}, {Ahn}, {Anderson}, {Aubourg}, {Bailey}, {Barkhouser}, {Bautista}, {Beifiori}, {Berlind}, {Bhardwaj}, {Bizyaev}, {Blake}, {Blanton}, {Blomqvist}, {Bolton}, {Borde}, {Bovy}, {Brandt}, {Brewington}, {Brinkmann}, {Brown}, {Brownstein}, {Bundy}, {Busca}, {Carithers}, {Carnero}, {Carr}, {Chen}, {Comparat}, {Connolly}, {Cope}, {Croft}, {Cuesta}, {da Costa}, {Davenport}, {Delubac}, {de Putter}, {Dhital}, {Ealet}, {Ebelke}, {Eisenstein}, {Escoffier}, {Fan}, {Filiz Ak}, {Finley}, {Font-Ribera}, {G{\'e}nova-Santos}, {Gunn}, {Guo}, {Haggard}, {Hall}, {Hamilton}, {Harris}, {Harris}, {Ho}, {Hogg}, {Holder}, {Honscheid}, {Huehnerhoff}, {Jordan}, {Jordan}, {Kauffmann}, {Kazin}, {Kirkby}, {Klaene}, {Kneib}, {Le Goff}, {Lee}, {Long}, {Loomis}, {Lundgren}, {Lupton}, {Maia}, {Makler}, {Malanushenko}, {Malanushenko}, {Mandelbaum}, {Manera}, {Maraston}, {Margala}, {Masters}, {McBride}, {McDonald}, {McGreer}, {McMahon}, {Mena}, {Miralda-Escud{\'e}}, {Montero-Dorta},
  {Montesano}, {Muna}, {Myers}, {Naugle}, {Nichol}, {Noterdaeme}, {Nuza}, {Olmstead}, {Oravetz}, {Oravetz}, {Owen}, {Padmanabhan}, {Palanque-Delabrouille}, {Pan}, {Parejko}, {P{\^a}ris}, {Percival}, {P{\'e}rez-Fournon}, {P{\'e}rez-R{\`a}fols}, {Petitjean}, {Pfaffenberger}, {Pforr}, {Pieri}, {Prada}, {Price-Whelan}, {Raddick}, {Rebolo}, {Rich}, {Richards}, {Rockosi}, {Roe}, {Ross}, {Ross}, {Rossi}, {Rubi{\~n}o-Martin}, {Samushia}, {S{\'a}nchez}, {Sayres}, {Schmidt}, {Schneider}, {Sc{\'o}ccola}, {Seo}, {Shelden}, {Sheldon}, {Shen}, {Shu}, {Slosar}, {Smee}, {Snedden}, {Stauffer}, {Steele}, {Strauss}, {Streblyanska}, {Suzuki}, {Swanson}, {Tal}, {Tanaka}, {Thomas}, {Tinker}, {Tojeiro}, {Tremonti}, {Vargas Maga{\~n}a}, {Verde}, {Viel}, {Wake}, {Watson}, {Weaver}, {Weinberg}, {Weiner}, {West}, {White}, {Wood-Vasey}, {Yeche}, {Zehavi}, {Zhao}, \& {Zheng}}]{dawson2013}
{Dawson}, K.~S., {Schlegel}, D.~J., {Ahn}, C.~P., {et~al.} 2013, \aj, 145, 10, \dodoi{10.1088/0004-6256/145/1/10}

\bibitem[{{Dietz} {et~al.}(2021){Dietz}, {Yoon}, {Beers}, {Placco}, \& {Lee}}]{dietz2021}
{Dietz}, S.~E., {Yoon}, J., {Beers}, T.~C., {Placco}, V.~M., \& {Lee}, Y.~S. 2021, \apj, 914, 100, \dodoi{10.3847/1538-4357/abefd6}

\bibitem[{{Frebel}(2018)}]{frebel2018}
{Frebel}, A. 2018, Annual Review of Nuclear and Particle Science, 68, 237, \dodoi{10.1146/annurev-nucl-101917-021141}

\bibitem[{{Frebel} {et~al.}(2006){Frebel}, {Christlieb}, {Norris}, {Beers}, {Bessell}, {Rhee}, {Fechner}, {Marsteller}, {Rossi}, {Thom}, {Wisotzki}, \& {Reimers}}]{frebel2006}
{Frebel}, A., {Christlieb}, N., {Norris}, J.~E., {et~al.} 2006, \apj, 652, 1585, \dodoi{10.1086/508506}

\bibitem[{{Gaia Collaboration} {et~al.}(2023){Gaia Collaboration}, {Vallenari}, {Brown}, {Prusti}, {de Bruijne}, {Arenou}, {Babusiaux}, {Biermann}, {Creevey}, {Ducourant}, {Evans}, {Eyer}, {Guerra}, {Hutton}, {Jordi}, {Klioner}, {Lammers}, {Lindegren}, {Luri}, {Mignard}, {Panem}, {Pourbaix}, {Randich}, {Sartoretti}, {Soubiran}, {Tanga}, {Walton}, {Bailer-Jones}, {Bastian}, {Drimmel}, {Jansen}, {Katz}, {Lattanzi}, {van Leeuwen}, {Bakker}, {Cacciari}, {Casta{\~n}eda}, {De Angeli}, {Fabricius}, {Fouesneau}, {Fr{\'e}mat}, {Galluccio}, {Guerrier}, {Heiter}, {Masana}, {Messineo}, {Mowlavi}, {Nicolas}, {Nienartowicz}, {Pailler}, {Panuzzo}, {Riclet}, {Roux}, {Seabroke}, {Sordo}, {Th{\'e}venin}, {Gracia-Abril}, {Portell}, {Teyssier}, {Altmann}, {Andrae}, {Audard}, {Bellas-Velidis}, {Benson}, {Berthier}, {Blomme}, {Burgess}, {Busonero}, {Busso}, {C{\'a}novas}, {Carry}, {Cellino}, {Cheek}, {Clementini}, {Damerdji}, {Davidson}, {de Teodoro}, {Nu{\~n}ez Campos}, {Delchambre}, {Dell'Oro}, {Esquej},
  {Fern{\'a}ndez-Hern{\'a}ndez}, {Fraile}, {Garabato}, {Garc{\'\i}a-Lario}, {Gosset}, {Haigron}, {Halbwachs}, {Hambly}, {Harrison}, {Hern{\'a}ndez}, {Hestroffer}, {Hodgkin}, {Holl}, {Jan{\ss}en}, {Jevardat de Fombelle}, {Jordan}, {Krone-Martins}, {Lanzafame}, {L{\"o}ffler}, {Marchal}, {Marrese}, {Moitinho}, {Muinonen}, {Osborne}, {Pancino}, {Pauwels}, {Recio-Blanco}, {Reyl{\'e}}, {Riello}, {Rimoldini}, {Roegiers}, {Rybizki}, {Sarro}, {Siopis}, {Smith}, {Sozzetti}, {Utrilla}, {van Leeuwen}, {Abbas}, {{\'A}brah{\'a}m}, {Abreu Aramburu}, {Aerts}, {Aguado}, {Ajaj}, {Aldea-Montero}, {Altavilla}, {{\'A}lvarez}, {Alves}, {Anders}, {Anderson}, {Anglada Varela}, {Antoja}, {Baines}, {Baker}, {Balaguer-N{\'u}{\~n}ez}, {Balbinot}, {Balog}, {Barache}, {Barbato}, {Barros}, {Barstow}, {Bartolom{\'e}}, {Bassilana}, {Bauchet}, {Becciani}, {Bellazzini}, {Berihuete}, {Bernet}, {Bertone}, {Bianchi}, {Binnenfeld}, {Blanco-Cuaresma}, {Blazere}, {Boch}, {Bombrun}, {Bossini}, {Bouquillon}, {Bragaglia}, {Bramante}, {Breedt},
  {Bressan}, {Brouillet}, {Brugaletta}, {Bucciarelli}, {Burlacu}, {Butkevich}, {Buzzi}, {Caffau}, {Cancelliere}, {Cantat-Gaudin}, {Carballo}, {Carlucci}, {Carnerero}, {Carrasco}, {Casamiquela}, {Castellani}, {Castro-Ginard}, {Chaoul}, {Charlot}, {Chemin}, {Chiaramida}, {Chiavassa}, {Chornay}, {Comoretto}, {Contursi}, {Cooper}, {Cornez}, {Cowell}, {Crifo}, {Cropper}, {Crosta}, {Crowley}, {Dafonte}, {Dapergolas}, {David}, {David}, {de Laverny}, {De Luise}, \& {De March}}]{gaia2023}
{Gaia Collaboration}, {Vallenari}, A., {Brown}, A.~G.~A., {et~al.} 2023, \aap, 674, A1, \dodoi{10.1051/0004-6361/202243940}

\bibitem[{{Gil-Pons} {et~al.}(2025){Gil-Pons}, {Campbell}, {Doherty}, \& {Lugaro}}]{gilpons2025}
{Gil-Pons}, P., {Campbell}, S.~W., {Doherty}, C.~L., \& {Lugaro}, M. 2025, arXiv e-prints, arXiv:2509.01302.
\newblock \doarXiv{2509.01302}

\bibitem[{{Hampel} {et~al.}(2016){Hampel}, {Stancliffe}, {Lugaro}, \& {Meyer}}]{hampel2016}
{Hampel}, M., {Stancliffe}, R.~J., {Lugaro}, M., \& {Meyer}, B.~S. 2016, \apj, 831, 171, \dodoi{10.3847/0004-637X/831/2/171}

\bibitem[{{Hansen} {et~al.}(2015){Hansen}, {Hansen}, {Christlieb}, {Beers}, {Yong}, {Bessell}, {Frebel}, {Garc{\'\i}a P{\'e}rez}, {Placco}, {Norris}, \& {Asplund}}]{hansen2015}
{Hansen}, T., {Hansen}, C.~J., {Christlieb}, N., {et~al.} 2015, \apj, 807, 173, \dodoi{10.1088/0004-637X/807/2/173}

\bibitem[{{Hansen} {et~al.}(2016{\natexlab{a}}){Hansen}, {Andersen}, {Nordstr{\"o}m}, {Beers}, {Placco}, {Yoon}, \& {Buchhave}}]{hansen2016a}
{Hansen}, T.~T., {Andersen}, J., {Nordstr{\"o}m}, B., {et~al.} 2016{\natexlab{a}}, \aap, 586, A160, \dodoi{10.1051/0004-6361/201527235}

\bibitem[{{Hansen} {et~al.}(2016{\natexlab{b}}){Hansen}, {Andersen}, {Nordstr{\"o}m}, {Beers}, {Placco}, {Yoon}, \& {Buchhave}}]{hansen2016b}
---. 2016{\natexlab{b}}, \aap, 588, A3, \dodoi{10.1051/0004-6361/201527409}

\bibitem[{{Hartwig} {et~al.}(2018){Hartwig}, {Yoshida}, {Magg}, {Frebel}, {Glover}, {G{\'o}mez}, {Griffen}, {Ishigaki}, {Ji}, {Klessen}, {O'Shea}, \& {Tominaga}}]{hartwig2018}
{Hartwig}, T., {Yoshida}, N., {Magg}, M., {et~al.} 2018, \mnras, 478, 1795, \dodoi{10.1093/mnras/sty1176}

\bibitem[{{Heger} \& {Woosley}(2010)}]{heger2010}
{Heger}, A., \& {Woosley}, S.~E. 2010, \apj, 724, 341, \dodoi{10.1088/0004-637X/724/1/341}

\bibitem[{{Herwig}(2005)}]{herwig2005}
{Herwig}, F. 2005, \araa, 43, 435, \dodoi{10.1146/annurev.astro.43.072103.150600}

\bibitem[{{Howes} {et~al.}(2016){Howes}, {Asplund}, {Keller}, {Casey}, {Yong}, {Lind}, {Frebel}, {Hays}, {Alves-Brito}, {Bessell}, {Casagrande}, {Marino}, {Nataf}, {Owen}, {Da Costa}, {Schmidt}, \& {Tisserand}}]{howes2016}
{Howes}, L.~M., {Asplund}, M., {Keller}, S.~C., {et~al.} 2016, \mnras, 460, 884, \dodoi{10.1093/mnras/stw1004}

\bibitem[{{Ishigaki} {et~al.}(2014){Ishigaki}, {Tominaga}, {Kobayashi}, \& {Nomoto}}]{ishigaki2014}
{Ishigaki}, M.~N., {Tominaga}, N., {Kobayashi}, C., \& {Nomoto}, K. 2014, \apjl, 792, L32, \dodoi{10.1088/2041-8205/792/2/L32}

\bibitem[{{Ito} {et~al.}(2013){Ito}, {Aoki}, {Beers}, {Tominaga}, {Honda}, \& {Carollo}}]{ito2013}
{Ito}, H., {Aoki}, W., {Beers}, T.~C., {et~al.} 2013, \apj, 773, 33, \dodoi{10.1088/0004-637X/773/1/33}

\bibitem[{{Ito} {et~al.}(2009){Ito}, {Aoki}, {Honda}, \& {Beers}}]{ito2009}
{Ito}, H., {Aoki}, W., {Honda}, S., \& {Beers}, T.~C. 2009, \apjl, 698, L37, \dodoi{10.1088/0004-637X/698/1/L37}

\bibitem[{{Iwamoto} {et~al.}(2005){Iwamoto}, {Umeda}, {Tominaga}, {Nomoto}, \& {Maeda}}]{iwamoto2005}
{Iwamoto}, N., {Umeda}, H., {Tominaga}, N., {Nomoto}, K., \& {Maeda}, K. 2005, Science, 309, 451, \dodoi{10.1126/science.1112997}

\bibitem[{{Jeon} {et~al.}(2021){Jeon}, {Bromm}, {Besla}, {Yoon}, \& {Choi}}]{jeon2021}
{Jeon}, M., {Bromm}, V., {Besla}, G., {Yoon}, J., \& {Choi}, Y. 2021, \mnras, 502, 1, \dodoi{10.1093/mnras/staa4017}

\bibitem[{{Jeong} {et~al.}(2023){Jeong}, {Lee}, {Beers}, {Placco}, {Kim}, {Koo}, {Lee}, \& {Yang}}]{jeong2023}
{Jeong}, M., {Lee}, Y.~S., {Beers}, T.~C., {et~al.} 2023, \apj, 948, 38, \dodoi{10.3847/1538-4357/acc58a}

\bibitem[{{Ji} {et~al.}(2020){Ji}, {Li}, {Simon}, {Marshall}, {Vivas}, {Pace}, {Bechtol}, {Drlica-Wagner}, {Koposov}, {Hansen}, {Allam}, {Gruendl}, {Johnson}, {McNanna}, {No{\"e}l}, {Tucker}, \& {Walker}}]{ji2020}
{Ji}, A.~P., {Li}, T.~S., {Simon}, J.~D., {et~al.} 2020, \apj, 889, 27, \dodoi{10.3847/1538-4357/ab6213}

\bibitem[{{Jorissen} {et~al.}(2016){Jorissen}, {Van Eck}, {Van Winckel}, {Merle}, {Boffin}, {Andersen}, {Nordstr{\"o}m}, {Udry}, {Masseron}, {Lenaerts}, \& {Waelkens}}]{jorissen2016}
{Jorissen}, A., {Van Eck}, S., {Van Winckel}, H., {et~al.} 2016, \aap, 586, A158, \dodoi{10.1051/0004-6361/201526992}

\bibitem[{{Komiya} {et~al.}(2007){Komiya}, {Suda}, {Minaguchi}, {Shigeyama}, {Aoki}, \& {Fujimoto}}]{komiya2007}
{Komiya}, Y., {Suda}, T., {Minaguchi}, H., {et~al.} 2007, \apj, 658, 367, \dodoi{10.1086/510826}

\bibitem[{{Komiya} {et~al.}(2020){Komiya}, {Suda}, {Yamada}, \& {Fujimoto}}]{komiya2020}
{Komiya}, Y., {Suda}, T., {Yamada}, S., \& {Fujimoto}, M.~Y. 2020, \apj, 890, 66, \dodoi{10.3847/1538-4357/ab67be}

\bibitem[{{Lau} {et~al.}(2007){Lau}, {Stancliffe}, \& {Tout}}]{lau2007}
{Lau}, H. H.~B., {Stancliffe}, R.~J., \& {Tout}, C.~A. 2007, \mnras, 378, 563, \dodoi{10.1111/j.1365-2966.2007.11773.x}

\bibitem[{{Lee} {et~al.}(2023){Lee}, {Lee}, {Kim}, {Beers}, \& {An}}]{lee2023}
{Lee}, A., {Lee}, Y.~S., {Kim}, Y.~K., {Beers}, T.~C., \& {An}, D. 2023, \apj, 945, 56, \dodoi{10.3847/1538-4357/acb6f5}

\bibitem[{{Lee} {et~al.}(2019){Lee}, {Beers}, \& {Kim}}]{lee2019}
{Lee}, Y.~S., {Beers}, T.~C., \& {Kim}, Y.~K. 2019, \apj, 885, 102, \dodoi{10.3847/1538-4357/ab4791}

\bibitem[{{Lee} {et~al.}(2017){Lee}, {Beers}, {Kim}, {Placco}, {Yoon}, {Carollo}, {Masseron}, \& {Jung}}]{lee2017}
{Lee}, Y.~S., {Beers}, T.~C., {Kim}, Y.~K., {et~al.} 2017, \apj, 836, 91, \dodoi{10.3847/1538-4357/836/1/91}

\bibitem[{{Lee} {et~al.}(2008{\natexlab{a}}){Lee}, {Beers}, {Sivarani}, {Johnson}, {An}, {Wilhelm}, {Allende Prieto}, {Koesterke}, {Re Fiorentin}, {Bailer-Jones}, {Norris}, {Yanny}, {Rockosi}, {Newberg}, {Cudworth}, \& {Pan}}]{lee2008a}
{Lee}, Y.~S., {Beers}, T.~C., {Sivarani}, T., {et~al.} 2008{\natexlab{a}}, \aj, 136, 2050, \dodoi{10.1088/0004-6256/136/5/2050}

\bibitem[{{Lee} {et~al.}(2008{\natexlab{b}}){Lee}, {Beers}, {Sivarani}, {Allende Prieto}, {Koesterke}, {Wilhelm}, {Re Fiorentin}, {Bailer-Jones}, {Norris}, {Rockosi}, {Yanny}, {Newberg}, {Covey}, {Zhang}, \& {Luo}}]{lee2008b}
---. 2008{\natexlab{b}}, \aj, 136, 2022, \dodoi{10.1088/0004-6256/136/5/2022}

\bibitem[{{Lee} {et~al.}(2011){Lee}, {Beers}, {An}, {Ivezi{\'c}}, {Just}, {Rockosi}, {Morrison}, {Johnson}, {Sch{\"o}nrich}, {Bird}, {Yanny}, {Harding}, \& {Rocha-Pinto}}]{lee2011}
{Lee}, Y.~S., {Beers}, T.~C., {An}, D., {et~al.} 2011, \apj, 738, 187, \dodoi{10.1088/0004-637X/738/2/187}

\bibitem[{{Lee} {et~al.}(2013){Lee}, {Beers}, {Masseron}, {Plez}, {Rockosi}, {Sobeck}, {Yanny}, {Lucatello}, {Sivarani}, {Placco}, \& {Carollo}}]{lee2013}
{Lee}, Y.~S., {Beers}, T.~C., {Masseron}, T., {et~al.} 2013, \aj, 146, 132, \dodoi{10.1088/0004-6256/146/5/132}

\bibitem[{{Lee} {et~al.}(2015){Lee}, {Beers}, {Carlin}, {Newberg}, {Hou}, {Li}, {Luo}, {Wu}, {Yang}, {Zhang}, {Zhang}, \& {Zhang}}]{lee2015}
{Lee}, Y.~S., {Beers}, T.~C., {Carlin}, J.~L., {et~al.} 2015, \aj, 150, 187, \dodoi{10.1088/0004-6256/150/6/187}

\bibitem[{{Li} {et~al.}(2022){Li}, {Aoki}, {Matsuno}, {Xing}, {Suda}, {Tominaga}, {Chen}, {Honda}, {Ishigaki}, {Shi}, {Zhao}, \& {Zhao}}]{li2022}
{Li}, H., {Aoki}, W., {Matsuno}, T., {et~al.} 2022, \apj, 931, 147, \dodoi{10.3847/1538-4357/ac6514}

\bibitem[{{Limongi} \& {Chieffi}(2018)}]{limongi2018}
{Limongi}, M., \& {Chieffi}, A. 2018, \apjs, 237, 13, \dodoi{10.3847/1538-4365/aacb24}

\bibitem[{{Liu} {et~al.}(2021){Liu}, {Sibony}, {Meynet}, \& {Bromm}}]{liu2021}
{Liu}, B., {Sibony}, Y., {Meynet}, G., \& {Bromm}, V. 2021, \mnras, 506, 5247, \dodoi{10.1093/mnras/stab2057}

\bibitem[{{Lucatello} {et~al.}(2006){Lucatello}, {Beers}, {Christlieb}, {Barklem}, {Rossi}, {Marsteller}, {Sivarani}, \& {Lee}}]{lucatello2006}
{Lucatello}, S., {Beers}, T.~C., {Christlieb}, N., {et~al.} 2006, \apjl, 652, L37, \dodoi{10.1086/509780}

\bibitem[{{Lucatello} {et~al.}(2003){Lucatello}, {Gratton}, {Cohen}, {Beers}, {Christlieb}, {Carretta}, \& {Ram{\'\i}rez}}]{lucatello2003}
{Lucatello}, S., {Gratton}, R., {Cohen}, J.~G., {et~al.} 2003, \aj, 125, 875, \dodoi{10.1086/345886}

\bibitem[{{Lucatello} {et~al.}(2005){Lucatello}, {Tsangarides}, {Beers}, {Carretta}, {Gratton}, \& {Ryan}}]{lucatello2005}
{Lucatello}, S., {Tsangarides}, S., {Beers}, T.~C., {et~al.} 2005, \apj, 625, 825, \dodoi{10.1086/428104}

\bibitem[{{Lucey} {et~al.}(2023){Lucey}, {Al Kharusi}, {Hawkins}, {Ting}, {Ramachandra}, {Price-Whelan}, {Beers}, {Lee}, \& {Yoon}}]{lucey2023}
{Lucey}, M., {Al Kharusi}, N., {Hawkins}, K., {et~al.} 2023, \mnras, 523, 4049, \dodoi{10.1093/mnras/stad1675}

\bibitem[{{Luo} {et~al.}(2015){Luo}, {Zhao}, {Zhao}, {Deng}, {Liu}, {Jing}, {Wang}, {Zhang}, {Shi}, {Cui}, {Chu}, {Li}, {Bai}, {Wu}, {Cai}, {Cao}, {Cao}, {Carlin}, {Chen}, {Chen}, {Chen}, {Chen}, {Chen}, {Chen}, {Chen}, {Christlieb}, {Chu}, {Cui}, {Dong}, {Du}, {Fan}, {Feng}, {Fu}, {Gao}, {Gong}, {Gu}, {Guo}, {Han}, {He}, {Hou}, {Hou}, {Hou}, {Hu}, {Hu}, {Hu}, {Huo}, {Jia}, {Jiang}, {Jiang}, {Jiang}, {Jin}, {Kong}, {Kong}, {Lei}, {Li}, {Li}, {Li}, {Li}, {Li}, {Li}, {Li}, {Li}, {Li}, {Li}, {Li}, {Li}, {Liang}, {Lin}, {Liu}, {Liu}, {Liu}, {Liu}, {Lu}, {Luo}, {Mao}, {Newberg}, {Ni}, {Qi}, {Qi}, {Shen}, {Shi}, {Song}, {Song}, {Su}, {Su}, {Tang}, {Tao}, {Tian}, {Wang}, {Wang}, {Wang}, {Wang}, {Wang}, {Wang}, {Wang}, {Wang}, {Wang}, {Wang}, {Wang}, {Wang}, {Wang}, {Wang}, {Wang}, {Wang}, {Wang}, {Wang}, {Wang}, {Wang}, {Wei}, {Wei}, {Wu}, {Wu}, {Wu}, {Wu}, {Xing}, {Xu}, {Xu}, {Xu}, {Yan}, {Yang}, {Yang}, {Yang}, {Yang}, {Yao}, {Yu}, {Yuan}, {Yuan}, {Yuan}, {Yuan}, {Zhai}, {Zhang}, {Zhang}, {Zhang}, {Zhang},
  {Zhang}, {Zhang}, {Zhang}, {Zhang}, {Zhao}, {Zhou}, {Zhou}, {Zhu}, {Zhu}, {Zou}, \& {Zuo}}]{luo2015}
{Luo}, A.~L., {Zhao}, Y.-H., {Zhao}, G., {et~al.} 2015, Research in Astronomy and Astrophysics, 15, 1095, \dodoi{10.1088/1674-4527/15/8/002}

\bibitem[{{Masseron} {et~al.}(2010){Masseron}, {Johnson}, {Plez}, {van Eck}, {Primas}, {Goriely}, \& {Jorissen}}]{masseron2010}
{Masseron}, T., {Johnson}, J.~A., {Plez}, B., {et~al.} 2010, \aap, 509, A93, \dodoi{10.1051/0004-6361/200911744}

\bibitem[{{Meynet} {et~al.}(2006){Meynet}, {Ekstr{\"o}m}, \& {Maeder}}]{meynet2006}
{Meynet}, G., {Ekstr{\"o}m}, S., \& {Maeder}, A. 2006, \aap, 447, 623, \dodoi{10.1051/0004-6361:20053070}

\bibitem[{{Meynet} {et~al.}(2010){Meynet}, {Hirschi}, {Ekstrom}, {Maeder}, {Georgy}, {Eggenberger}, \& {Chiappini}}]{meynet2010}
{Meynet}, G., {Hirschi}, R., {Ekstrom}, S., {et~al.} 2010, \aap, 521, A30, \dodoi{10.1051/0004-6361/200913377}

\bibitem[{{Molaro} {et~al.}(2023){Molaro}, {Aguado}, {Caffau}, {Allende Prieto}, {Bonifacio}, {Gonz{\'a}lez Hern{\'a}ndez}, {Rebolo}, {Zapatero Osorio}, {Cristiani}, {Pepe}, {Santos}, {Alibert}, {Cupani}, {Di Marcantonio}, {D'Odorico}, {Lovis}, {Martins}, {Milakovi{\'c}}, {Murphy}, {Nunes}, {Schmidt}, {Sousa}, {Sozzetti}, \& {Su{\'a}rez Mascare{\~n}o}}]{molaro2023}
{Molaro}, P., {Aguado}, D.~S., {Caffau}, E., {et~al.} 2023, \aap, 679, A72, \dodoi{10.1051/0004-6361/202347676}

\bibitem[{{Nomoto} {et~al.}(2013){Nomoto}, {Kobayashi}, \& {Tominaga}}]{nomoto2013}
{Nomoto}, K., {Kobayashi}, C., \& {Tominaga}, N. 2013, \araa, 51, 457, \dodoi{10.1146/annurev-astro-082812-140956}

\bibitem[{{Norris} {et~al.}(1997){Norris}, {Ryan}, \& {Beers}}]{norris1997}
{Norris}, J.~E., {Ryan}, S.~G., \& {Beers}, T.~C. 1997, \apjl, 489, L169, \dodoi{10.1086/316787}

\bibitem[{{Norris} {et~al.}(2013){Norris}, {Yong}, {Bessell}, {Christlieb}, {Asplund}, {Gilmore}, {Wyse}, {Beers}, {Barklem}, {Frebel}, \& {Ryan}}]{norris2013}
{Norris}, J.~E., {Yong}, D., {Bessell}, M.~S., {et~al.} 2013, \apj, 762, 28, \dodoi{10.1088/0004-637X/762/1/28}

\bibitem[{{Placco} {et~al.}(2014){Placco}, {Frebel}, {Beers}, \& {Stancliffe}}]{placco2014}
{Placco}, V.~M., {Frebel}, A., {Beers}, T.~C., \& {Stancliffe}, R.~J. 2014, \apj, 797, 21, \dodoi{10.1088/0004-637X/797/1/21}

\bibitem[{{Rockosi} {et~al.}(2022){Rockosi}, {Lee}, {Morrison}, {Yanny}, {Johnson}, {Lucatello}, {Sobeck}, {Beers}, {Allende Prieto}, {An}, {Bizyaev}, {Blanton}, {Casagrande}, {Eisenstein}, {Gould}, {Gunn}, {Harding}, {Ivans}, {Jacobson}, {Janesh}, {Knapp}, {Kollmeier}, {L{\'e}pine}, {L{\'o}pez-Corredoira}, {Ma}, {Newberg}, {Pan}, {Prchlik}, {Sayers}, {Schlesinger}, {Simmerer}, \& {Weinberg}}]{rockosi2022}
{Rockosi}, C.~M., {Lee}, Y.~S., {Morrison}, H.~L., {et~al.} 2022, \apjs, 259, 60, \dodoi{10.3847/1538-4365/ac5323}

\bibitem[{{Rossi} {et~al.}(2023){Rossi}, {Salvadori}, {Sk{\'u}lad{\'o}ttir}, \& {Vanni}}]{rossi2023}
{Rossi}, M., {Salvadori}, S., {Sk{\'u}lad{\'o}ttir}, {\'A}., \& {Vanni}, I. 2023, \mnras, 522, L1, \dodoi{10.1093/mnrasl/slad029}

\bibitem[{{Rossi} {et~al.}(2024){Rossi}, {Salvadori}, {Sk{\'u}lad{\'o}ttir}, {Vanni}, \& {Koutsouridou}}]{rossi2024}
{Rossi}, M., {Salvadori}, S., {Sk{\'u}lad{\'o}ttir}, {\'A}., {Vanni}, I., \& {Koutsouridou}, I. 2024, arXiv e-prints, arXiv:2406.12960, \dodoi{10.48550/arXiv.2406.12960}

\bibitem[{{Rossi} {et~al.}(2005){Rossi}, {Beers}, {Sneden}, {Sevastyanenko}, {Rhee}, \& {Marsteller}}]{rossi2005}
{Rossi}, S., {Beers}, T.~C., {Sneden}, C., {et~al.} 2005, \aj, 130, 2804, \dodoi{10.1086/497164}

\bibitem[{{Salvadori} {et~al.}(2015){Salvadori}, {Sk{\'u}lad{\'o}ttir}, \& {Tolstoy}}]{salvadori2015}
{Salvadori}, S., {Sk{\'u}lad{\'o}ttir}, {\'A}., \& {Tolstoy}, E. 2015, \mnras, 454, 1320, \dodoi{10.1093/mnras/stv1969}

\bibitem[{{Sharma} {et~al.}(2019){Sharma}, {Theuns}, \& {Frenk}}]{sharma2019}
{Sharma}, M., {Theuns}, T., \& {Frenk}, C. 2019, \mnras, 482, L145, \dodoi{10.1093/mnrasl/sly195}

\bibitem[{{Sivarani} {et~al.}(2006){Sivarani}, {Beers}, {Bonifacio}, {Molaro}, {Cayrel}, {Herwig}, {Spite}, {Spite}, {Plez}, {Andersen}, {Barbuy}, {Depagne}, {Hill}, {Fran{\c{c}}ois}, {Nordstr{\"o}m}, \& {Primas}}]{sivarani2006}
{Sivarani}, T., {Beers}, T.~C., {Bonifacio}, P., {et~al.} 2006, \aap, 459, 125, \dodoi{10.1051/0004-6361:20065440}

\bibitem[{{Smolinski} {et~al.}(2011){Smolinski}, {Martell}, {Beers}, \& {Lee}}]{smolinski2011}
{Smolinski}, J.~P., {Martell}, S.~L., {Beers}, T.~C., \& {Lee}, Y.~S. 2011, \aj, 142, 126, \dodoi{10.1088/0004-6256/142/4/126}

\bibitem[{{Sneden} {et~al.}(2008){Sneden}, {Cowan}, \& {Gallino}}]{sneden2008}
{Sneden}, C., {Cowan}, J.~J., \& {Gallino}, R. 2008, \araa, 46, 241, \dodoi{10.1146/annurev.astro.46.060407.145207}

\bibitem[{{Spite} {et~al.}(2013){Spite}, {Caffau}, {Bonifacio}, {Spite}, {Ludwig}, {Plez}, \& {Christlieb}}]{spite2013}
{Spite}, M., {Caffau}, E., {Bonifacio}, P., {et~al.} 2013, \aap, 552, A107, \dodoi{10.1051/0004-6361/201220989}

\bibitem[{{Stancliffe} {et~al.}(2009){Stancliffe}, {Church}, {Angelou}, \& {Lattanzio}}]{stancliffe2009}
{Stancliffe}, R.~J., {Church}, R.~P., {Angelou}, G.~C., \& {Lattanzio}, J.~C. 2009, \mnras, 396, 2313, \dodoi{10.1111/j.1365-2966.2009.14900.x}

\bibitem[{{Starkenburg} {et~al.}(2014){Starkenburg}, {Shetrone}, {McConnachie}, \& {Venn}}]{starkenburg2014}
{Starkenburg}, E., {Shetrone}, M.~D., {McConnachie}, A.~W., \& {Venn}, K.~A. 2014, \mnras, 441, 1217, \dodoi{10.1093/mnras/stu623}

\bibitem[{{Starkenburg} {et~al.}(2013){Starkenburg}, {Hill}, {Tolstoy}, {Fran{\c{c}}ois}, {Irwin}, {Boschman}, {Venn}, {de Boer}, {Lemasle}, {Jablonka}, {Battaglia}, {Groot}, \& {Kaper}}]{starkenburg2013}
{Starkenburg}, E., {Hill}, V., {Tolstoy}, E., {et~al.} 2013, \aap, 549, A88, \dodoi{10.1051/0004-6361/201220349}

\bibitem[{{Suda} {et~al.}(2004){Suda}, {Aikawa}, {Machida}, {Fujimoto}, \& {Iben}}]{suda2004}
{Suda}, T., {Aikawa}, M., {Machida}, M.~N., {Fujimoto}, M.~Y., \& {Iben}, Jr., I. 2004, \apj, 611, 476, \dodoi{10.1086/422135}

\bibitem[{{Suda} {et~al.}(2008){Suda}, {Katsuta}, {Yamada}, {Suwa}, {Ishizuka}, {Komiya}, {Sorai}, {Aikawa}, \& {Fujimoto}}]{suda2008}
{Suda}, T., {Katsuta}, Y., {Yamada}, S., {et~al.} 2008, \pasj, 60, 1159, \dodoi{10.1093/pasj/60.5.1159}

\bibitem[{{Tominaga} {et~al.}(2014){Tominaga}, {Iwamoto}, \& {Nomoto}}]{tominaga2014}
{Tominaga}, N., {Iwamoto}, N., \& {Nomoto}, K. 2014, \apj, 785, 98, \dodoi{10.1088/0004-637X/785/2/98}

\bibitem[{{Tominaga} {et~al.}(2007){Tominaga}, {Umeda}, \& {Nomoto}}]{tominaga2007}
{Tominaga}, N., {Umeda}, H., \& {Nomoto}, K. 2007, \apj, 660, 516, \dodoi{10.1086/513063}

\bibitem[{{Umeda} \& {Nomoto}(2003)}]{umeda2003}
{Umeda}, H., \& {Nomoto}, K. 2003, \nat, 422, 871, \dodoi{10.1038/nature01571}

\bibitem[{{Umeda} \& {Nomoto}(2005)}]{umeda2005}
---. 2005, \apj, 619, 427, \dodoi{10.1086/426097}

\bibitem[{Wilson(1927)}]{wilson1927}
Wilson, E.~B. 1927, Journal of the American Statistical Association, 22, 209, \dodoi{10.1080/01621459.1927.10502953}

\bibitem[{{Yanny} {et~al.}(2009){Yanny}, {Newberg}, {Johnson}, {Lee}, {Beers}, {Bizyaev}, {Brewington}, {Fiorentin}, {Harding}, {Malanushenko}, {Malanushenko}, {Oravetz}, {Pan}, {Simmons}, \& {Snedden}}]{yanny2009}
{Yanny}, B., {Newberg}, H.~J., {Johnson}, J.~A., {et~al.} 2009, \apj, 700, 1282, \dodoi{10.1088/0004-637X/700/2/1282}

\bibitem[{{Yong} {et~al.}(2013){Yong}, {Norris}, {Bessell}, {Christlieb}, {Asplund}, {Beers}, {Barklem}, {Frebel}, \& {Ryan}}]{yong2013}
{Yong}, D., {Norris}, J.~E., {Bessell}, M.~S., {et~al.} 2013, \apj, 762, 27, \dodoi{10.1088/0004-637X/762/1/27}

\bibitem[{{Yoon} {et~al.}(2019){Yoon}, {Beers}, {Tian}, \& {Whitten}}]{yoon2019}
{Yoon}, J., {Beers}, T.~C., {Tian}, D., \& {Whitten}, D.~D. 2019, \apj, 878, 97, \dodoi{10.3847/1538-4357/ab1ead}

\bibitem[{{Yoon} {et~al.}(2016){Yoon}, {Beers}, {Placco}, {Rasmussen}, {Carollo}, {He}, {Hansen}, {Roederer}, \& {Zeanah}}]{yoon2016}
{Yoon}, J., {Beers}, T.~C., {Placco}, V.~M., {et~al.} 2016, \apj, 833, 20, \dodoi{10.3847/0004-637X/833/1/20}

\bibitem[{{Yoon} {et~al.}(2018){Yoon}, {Beers}, {Dietz}, {Lee}, {Placco}, {Da Costa}, {Keller}, {Owen}, \& {Sharma}}]{yoon2018}
{Yoon}, J., {Beers}, T.~C., {Dietz}, S., {et~al.} 2018, \apj, 861, 146, \dodoi{10.3847/1538-4357/aaccea}

\bibitem[{{York} {et~al.}(2000){York}, {Adelman}, {Anderson}, {Anderson}, {Annis}, {Bahcall}, {Bakken}, {Barkhouser}, {Bastian}, {Berman}, {Boroski}, {Bracker}, {Briegel}, {Briggs}, {Brinkmann}, {Brunner}, {Burles}, {Carey}, {Carr}, {Castander}, {Chen}, {Colestock}, {Connolly}, {Crocker}, {Csabai}, {Czarapata}, {Davis}, {Doi}, {Dombeck}, {Eisenstein}, {Ellman}, {Elms}, {Evans}, {Fan}, {Federwitz}, {Fiscelli}, {Friedman}, {Frieman}, {Fukugita}, {Gillespie}, {Gunn}, {Gurbani}, {de Haas}, {Haldeman}, {Harris}, {Hayes}, {Heckman}, {Hennessy}, {Hindsley}, {Holm}, {Holmgren}, {Huang}, {Hull}, {Husby}, {Ichikawa}, {Ichikawa}, {Ivezi{\'c}}, {Kent}, {Kim}, {Kinney}, {Klaene}, {Kleinman}, {Kleinman}, {Knapp}, {Korienek}, {Kron}, {Kunszt}, {Lamb}, {Lee}, {Leger}, {Limmongkol}, {Lindenmeyer}, {Long}, {Loomis}, {Loveday}, {Lucinio}, {Lupton}, {MacKinnon}, {Mannery}, {Mantsch}, {Margon}, {McGehee}, {McKay}, {Meiksin}, {Merelli}, {Monet}, {Munn}, {Narayanan}, {Nash}, {Neilsen}, {Neswold}, {Newberg}, {Nichol}, {Nicinski},
  {Nonino}, {Okada}, {Okamura}, {Ostriker}, {Owen}, {Pauls}, {Peoples}, {Peterson}, {Petravick}, {Pier}, {Pope}, {Pordes}, {Prosapio}, {Rechenmacher}, {Quinn}, {Richards}, {Richmond}, {Rivetta}, {Rockosi}, {Ruthmansdorfer}, {Sandford}, {Schlegel}, {Schneider}, {Sekiguchi}, {Sergey}, {Shimasaku}, {Siegmund}, {Smee}, {Smith}, {Snedden}, {Stone}, {Stoughton}, {Strauss}, {Stubbs}, {SubbaRao}, {Szalay}, {Szapudi}, {Szokoly}, {Thakar}, {Tremonti}, {Tucker}, {Uomoto}, {Vanden Berk}, {Vogeley}, {Waddell}, {Wang}, {Watanabe}, {Weinberg}, {Yanny}, {Yasuda}, \& {SDSS Collaboration}}]{york2000}
{York}, D.~G., {Adelman}, J., {Anderson}, Jr., J.~E., {et~al.} 2000, \aj, 120, 1579, \dodoi{10.1086/301513}

\end{thebibliography}
\end{CJK}
\end{document}